\documentclass[aps,prd,preprint,showpacs,superscriptaddress,groupedaddress,nofootinbib,tightenlines
]{revtex4}
\usepackage{amsmath, amsfonts, amsthm, amssymb, ulem}
\usepackage{epsfig}
\usepackage{subfigure}
\usepackage{float}
\usepackage{verbatim} 
\usepackage{color}
\usepackage{slashed}
\usepackage{multirow}
\usepackage{booktabs}
\usepackage{epstopdf}

\allowdisplaybreaks[3]

\def\({\left(}     \def\){\right)}     \def\[{\left[}    \def\]{\right]}   \def\d{\mathrm{d}}
     
\newcommand{\f}[2]{\frac{#1}{#2}}
\def \bal#1\eal  {\begin{align} #1 \end{align}}
\newcommand{\be} {\begin{equation}}    \newcommand{\ee} {\end{equation}}
\newcommand{\bc}{\begin{center}}   \newcommand{\ec}{\end{center}}
\newcommand{\bim} {\begin{itemize}[noitemsep]}      \newcommand{\eim} {\end{itemize}}

\newcommand{\nn} {\nonumber\\}
     
\newcommand{\ud} {\mathrm{d}}    
  \newcommand{\pd} {\partial}  

\newcommand{\ai}{{\alpha}}   \newcommand{\bi}{{\beta}}   \newcommand{\gi}{{\gamma}}
        
\newcommand{\li}{{\lambda}}      
\newcommand{\oi}{\omega}

\begin{document}

\rightline {\small USTC-ICTS-1932}

\title{Oscillon Preheating in Full General Relativity}

\author{Xiao-Xiao Kou}
\email{kxx1998@mail.ustc.edu.cn}

\affiliation{Interdisciplinary Center for Theoretical Study, University of Science and Technology of China, Hefei, Anhui 230026, China}
\affiliation{Peng Huanwu Center for Fundamental Theory, Hefei, Anhui 230026, China}

\author{Chi Tian}
\email{chi.tian@case.edu}
\affiliation{CERCA/ISO, Department of Physics, Case Western Reserve University, 10900, Euclid Avenue, Cleveland, OH 44106, USA
}

\author{Shuang-Yong Zhou}
\email{zhoushy@ustc.edu.cn}
\affiliation{Interdisciplinary Center for Theoretical Study, University of Science and Technology of China, Hefei, Anhui 230026, China}
\affiliation{Peng Huanwu Center for Fundamental Theory, Hefei, Anhui 230026, China}

\begin{abstract}
  
Oscillons are dense nonperturbative objects that may be copiously produced in the preheating period after inflation. Oscillon preheatings are usually simulated with nonlinear matter interactions but in a rigid FLRW background, without taking into account the gravitational backreactions. We investigate the oscillon preheating scenario in full general relativity with a full numerical relativity scheme, and compare the general-relativistic simulations with the traditional ones done in an FLRW background. We find that in certain parameter regions the general-relativistic corrections are significant, producing more and denser oscillons, and can be strong enough to collapse the oscillons to black holes. 

\end{abstract}

\maketitle
\flushbottom

\section{Introduction}

Oscillons are oscillating quasi-solitons that are spatially localized and have a long lifetime in a class of scalar field theories whose potentials can generate attractive forces between inhomogeneities \cite{Bogolyubsky:1976yu, Gleiser:1993pt, Copeland:1995fq, Honda:2001xg, Hindmarsh:2006ur, Fodor:2006zs, Saffin:2006yk, Hindmarsh:2007jb, Fodor:2008es, Gleiser:2008ty, Hertzberg:2010yz, Salmi:2012ta, Amin:2013ika, Copeland:2014qra, Krippendorf:2018tei, Amin:2019ums}. Thanks to advances in numerical tools and methods, their fascinating nonlinear properties and dynamics have been extensively investigated.  Oscillons may play an important role in the early universe \cite{Gleiser:2006te, Graham:2006xs, Amin:2010xe, Amin:2010jq}, particularly in the period of preheating \cite{Amin:2011hj, Amin:2010dc, Broadhead:2005hn, Farhi:2007wj, Gleiser:2011xj, Zhou:2013tsa, Antusch:2015ziz, Lozanov:2019ylm}, when particles are explosively generated as inflation ends and parametric resonance takes place (see \cite{Traschen:1990sw,Dolgov:1989sy,Shtanov:1994ce,Khlebnikov:1996mc,Kofman:1997yn,Felder:2000hj} for pioneering works in preheating). In the oscillon preheating scenario, the inflaton condensate fragments into oscillons during the preheating process, leading to a period of oscillon dominated universe whose expansion rate mimics the matter dominated universe \cite{Amin:2011hj}. A stochastic background of gravitational waves can be produced during oscillon preheating \cite{Zhou:2013tsa,  Antusch:2016con, Liu:2017hua, Antusch:2017flz, Antusch:2017vga, Zhou:2015yfa, Amin:2018xfe, Sang:2019ndv}. If the energy scale of the oscillon preheating or preheating-like period is sufficiently low, it may lead to interesting observable signals in the aLIGO experiments \cite{Liu:2017hua}.
On the other hand, scalar theories that support oscillons are those whose potentials are shallower than the quadratic ones, and, indeed, recent cosmological observations favor an inflationary potential that is flatter than the quadratic potential \cite{Aghanim:2018eyx}.

The dynamics of preheating has been traditionally studied with nonlinear matter simulations in the homogeneous FLRW background driven by the spatially averaged matter fields. This approach only keeps the homogeneous mode of the metric but ignores the inhomogeneous gravitational backreaction. However, oscillons are densely concentrated objects, which prompts the question whether the backreaction will become important in oscillon preheating at least in some parameter space. To this end, numerical simulations with full general-relativistic effects are needed, which account for all the non-linear and non-perturbative strong gravitational effects by solving the Einstein equations directly. Seeking full general-relativistic solutions in cosmology has been explored in pre-inflationary era \cite{East:2015ggf, East:2016anr, Clough:2017efm, Bloomfield:2019rbs} and in the late universe scenarios \cite{Giblin:2015vwq, Bentivegna:2015flc, Giblin:2016mjp, Macpherson:2016ict, Giblin:2017juu, East:2017qmk, Wang:2018qfr, Macpherson:2018akp, Macpherson:2018btl, Giblin:2018ndw, Muia:2019coe}. 
The role of general relativity in a preheating model without oscillon formation has been very recently discussed in \cite{Giblin:2019nuv}.

In this paper, we will perform the first study on oscillon preheating with the full power of numerical relativity, and compare the results to the simulations in an FLRW background. As we will see, for some parameter space, general-relativistic effects can be manifest, which tend to condensate more and denser oscillons than the traditional treatment with an FLRW background, and formation of primordial black holes can also be identified.

The paper is organized as follows. In Section \ref{sec:2}, we introduce the model to be studied and the numerical relativity setup, including our evolution code, the initial and gauge conditions, the grid setting, the tests of our code and so on. In Section \ref{sec:1}, we present the results of oscillon preheating in general relativity and compare them to the results in the traditional FLRW simulations. In Section \ref{sec:amr}, we switch on the AMR functionality to resolve those oscillons where self-gravity is strong and thus more resolutions are needed. In Section \ref{sec:BH}, we show that those strong self-gravitating oscillons can collapse to black holes, and in Section \ref{sec:largeBH}, we show how the mass of these primordial black holes scales in low scale ``preheating'' models. We conclude in Section \ref{sec:clu}.

\section{Model and setup}
\label{sec:2}

Oscillons \cite{Bogolyubsky:1976yu, Gleiser:1993pt, Copeland:1995fq} are localized quasi-stationary field configurations that arise in scalar field theories with certain shallow potentials, traditionally neglecting the gravitational backreactions. They exist because the scalar potential is flatter than a free quadratic potential away from the minimum, and so in these field theories ``particles'' prefer to condensate to form a lamp rather than propagate away to dissipate. The simplest oscillon configuration roughly goes like $\phi\simeq f(r)\cos \oi t$, where $f(r)$ is a spherically symmetric profile and the oscillating frequency $\oi$ is slightly smaller than $m$, the particle mass around the background $\phi=0$. They are very much like their U(1) symmetric counterpart Q-balls \cite{Coleman:1985ki} where the field configuration goes like $\Phi=f(r)e^{i\oi t}$ with $\Phi$ being a complex scalar field\,\footnote{Composite quasi-stable Q-balls also exit and have more complex inner structures such as charge swapping within the ball \cite{Copeland:2014qra}.}. But unlike the Q-balls, oscillons are only quasi-stable and quasi-stationary, and thus they decay in finite times, and the temporal and spatial dependence of an oscillon does not exactly factorize like the elementary Q-balls. Nevertheless, oscillons are attractor solutions with broad basins of attraction and can form from quite generic initial conditions \cite{Andersen:2012wg}. During reheating after inflation, the inflaton condensate starts to oscillate and the field perturbations often undergo a process of parametric resonance, where every Fourier mode of the perturbations satisfies a modified Mathieu's equation. This kind of reheating scenario where parametric resonance takes place is called preheating \cite{Traschen:1990sw, Dolgov:1989us}. Indeed, in preheating particles or perturbations are generated very efficiently, and if the potential allows, oscillons can often be copiously generated in the reheating process \cite{Amin:2011hj}, which will be referred to as the oscillon preheating scenario. In other words, the favorable conditions in reheating after inflaton provide a platform where oscillons play a role in the early cosmic evolution.

While oscillon preheating arises in many scenarios where the inflation potential is sufficiently flat away from the (quadratic) minimum,  we will focus on a representative class of minimally coupled models given by the action
\be
\label{action00}
S=\int \ud^4 x \sqrt{-g} \( \frac{M_{\rm pl}^2}{2} R - \f12 \pd_\mu\phi\pd^\mu \phi - V(\phi) \) ,
\ee
with the potential
\begin{align}
\label{eq:3}
V(\phi) = \frac{m^2 M^2}{ 2\alpha} \left[\left(1+\frac{\phi^2}{M^2}\right)^{\alpha} - 1\right]  ,
\end{align}
which is parametrized by the mass of the inflaton $m$ and two dimensionless parameters $\alpha$ and $\beta \equiv M_{\rm pl}/M$ ($M_{\rm pl}$ being the reduced Planck mass).  
Apart from Sec.~\ref{sec:largeBH} where we discuss implications of black holes collapsed from oscillons in a low scale ``preheating'' scenario, we fix $m$ by matching to the power spectrum of curvature perturbations from most recent CMB observations \cite{Aghanim:2018eyx}
\begin{align}
\label{eq:7}
  A_s = \frac{(4\alpha {\cal N})^{1+\alpha}}{96\pi^2 \alpha^3} \left(\frac{m}{M_{\rm pl}}\right)^2 \!\!  \left(\frac{M}{M_{\rm pl}}\right)^{2-2\alpha} \!\! \simeq 2.1 \! \times\! 10^{-9} ,
\end{align}
where ${\cal N} \simeq 50$ is the e-folds to the end of inflation, and the range of $0<\alpha < 1$ and $0 < \beta < 100$ is considered. For $\alpha<1$, the potential is flatter than the quadratic mass term at large $\phi$ and interpolates to the mass term at small $\phi$, thus belonging to the ``open-up'' type and potentially supporting oscillon formations in the preheating period. It is interesting to notice that flatter potentials are favored observationally \cite{Aghanim:2018eyx}, and this particular class of models are also motivated by inflation model constructions in string/M theory \cite{Silverstein:2008sg, McAllister:2008hb, Dong:2010in}.

The metric satisfies Einstein's equations and the scalar field obeys the Klein-Gordon equation $\Box \phi = {\d V}/{\d \phi}$. In traditional oscillon preheating simulations, the metric is fixed to be the homogeneous FLRW form and the Klein-Gordon equation reduces to
\begin{align}
\label{eq:10}
\ddot{\phi} + 3 H \dot{\phi} - \frac{\nabla^2 \phi}{a^2} + \frac{{\rm d}V(\phi)}{\rm{d}\phi} = 0, 
\end{align}
where $a$ is the scale factor, and the time evolution of the Hubble parameter $H$ follows
the Friedmann equation that depends on the spatial averaging of the energy momentum tensor. We will refer to this as {\it the FLRW scheme}. In this scheme, the gravitational backreactions above the FLRW background are assumed to be negligible, which might be more or less justified for typical preheating scenarios \cite{Giblin:2019nuv}, but oscillons are localized objects with centralizing energy densities, so it is {\it a priori} unclear whether and when full general relativity effects can be neglected in oscillon preheating.

In the FLRW scheme, by a Floquet analysis, $\phi$ in momentum space goes approximately as 
\begin{align}
\label{eq:9}
\phi_k =P_+ (t) e^{\mu_k t} - P_-(t) e^{-\mu_k t},
\end{align}
and thus the formation rate of oscillons is linked to the real part of Floquet exponent $\Re(\mu_k)$. Oscillons can efficiently form when strong parametric resonance occurs in the preheating period, which is roughly when $|\Re(\mu_k)|/H \gtrsim 7$, and the maximum value of $|\Re(\mu_k)|/H$ is approximately proportional to $\beta$  \cite{Amin:2011hj}. Since in the initial period of preheating perturbations from all the fields are very small and oscillons are yet to form, it is expected that this analysis should carry mostly unchanged to the full general relativity case. 

As mentioned, the FLRW scheme, while simple to implement and useful for many purposes, may not fully resolve the dynamics of oscillons where self-gravity becomes strong, for which case we need a fully general-relativistic treatment (to be referred to as {\it the full GR scheme}). To this end, we follow the conventional 3+1 formalism and decompose the spacetime metric into the form
\begin{align}
\label{eq:6}
g_{\mu\nu} = 
\begin{bmatrix}
-N^2 + N_kN^k & N_j \\
N_i       & \gi_{ij}   \\
\end{bmatrix},
\end{align}
where $N$ and $N^i$ are the lapse function and shift vector respectively and $\gamma_{ij}$ is the spatial metric. We also cast the Klein-Gordon equation into a hyperbolic form
\begin{align}
\label{eq:5}
\dot{\phi} &= N^i \pd_i\phi -N \Pi
\\
  \dot{\Pi}   &=    N^k\pd_k \Pi - N \gi^{ij} \pd_i  \psi_j + N \gi^{ij} \Gamma^{k}_{ij}\psi_k  +  N K \Pi\nn
                &-   \gi^{ij} \psi_i \pd_j N  +  N {\rm d}V(\phi)/{\rm d}\phi \\
 \dot{\psi}_i &= N^j\pd_j \psi_i + \psi_j \pd_i N^j -N \pd_i \Pi -\Pi \pd_i N,
\end{align}
where $\psi_i$ is the auxiliary field $\psi_i \equiv \partial_i \phi$, $K$ is the trace of the extrinsic curvature and $\Pi$ is the canonical momentum of the scalar field
\begin{align}
\label{eq:8}
\Pi \equiv - \left(\frac{1}{N}\dot{\phi} - \frac{1}{N} N^k\partial_k\phi \right) .
\end{align}
To evolve the metric components together with the scalar field they couple to, we employ the grid-based numerical relativity code \textsc{CosmoGRaPH} \cite{Mertens:2015ttp}, which makes use of the popular Baumgarte--Shapiro--Shibata--Nakamura (BSSN) formalism \cite{Nakamura:1987zz,Shibata:1995we, Baumgarte:1998te} and integrates the Adaptive Mesh Refinement (AMR) framework into its spatial grid scheme. The \textsc{CosmoGRaPH} code has been used to investigate GR effects in various cosmological scenarios, from structure formations \cite{Giblin:2015vwq, Giblin:2016mjp, Giblin:2018ndw} to inhomogeneous cosmological models \cite{Giblin:2019pql}. Its AMR feature is based on SAMRAI \cite{DBLP:conf/sc/WissinkHKSE01}, an open-source AMR application infrastructure, which has been proven to be scalable with million-cores \cite{GUNNEY201665}. To identify black holes that might have formed during the evolution, we have imported the AHFinderDirect package \cite{Thornburg:2003sf}, integrated as a subroutine to detect any apparent horizon existing on the hypersurface.

To facilitate an easy comparison to the FLRW treatment, we seek exact solutions to the constraint equations that mostly resemble the FLRW metric. We follow the standard procedure of conformal decomposition by redefining $\gamma_{ij} \equiv \Psi^4 \tilde{\gamma}_{ij}$ and the extrinsic curvature $K_{ij} \equiv A_{ij}+\frac13 \gamma_{ij} K$, and we choose the initial ansatz such that $\tilde{\gamma}_{ij} = \delta_{ij}$, $A_{ij} = 0$. Since $K=-3H$, we also choose an initial homogeneous $K$ such that the corresponding Hubble parameter satisfies the Friedmann equation. The initial configuration of the scalar field is set to satisfy $\dot{\phi}=0$ and a standard spectrum of the initial vacuum fluctuations for $\phi=M$. We then solve the non-linear constraint equation for $\Psi$ by employing the multigrid constraint solver integrated within \textsc{CosmoGRaPH}, which utilizes a full multigrid iteration scheme and an inexact-Newton-relaxation method \cite{Press_2003}. To ensure numerical convergence of the elliptic constraint solver, a cut-off at wavenumber $k=8$ is implemented.

The gauge conditions used in our full GR simulation is a modified version
of the widely employed ``1+log'' and ``Gamma-driver'' conditions:
\begin{eqnarray}
\label{eq:gaugecondition}
  \partial_t N &=& -2 \eta N \left(K - \langle K \rangle\right) + N^i \partial_i N,  \\
  \partial_t N^i  & = & B^i, \;\;\;\;  \partial_t B^i  =  \frac34 \partial_t \tilde{\Gamma}^i - B^i, 
\end{eqnarray}
where $\eta$ is chosen to be $0.5$ and $\tilde{\Gamma}^i \! \equiv \! \tilde{\gamma}^{jk} \tilde{\Gamma}_{jk}^i$, $\tilde{\Gamma}_{jk}^i$ being the Christoffel symbols of $\tilde{\gamma}_{ij}$.
This combination of gauge choices has been shown to have powerful singularity-avoidance properties \cite{Brown:2007pg}.
For better numerical stability and to maximally mimic the behavior of the FLRW scheme, the ``1+log'' gauge used here is slightly different from the usual one by an extra reference expansion rate, $\langle K \rangle$, which is the conformal average of the extrinsic curvature $K$ over the whole spatial hypersurface. For both the FLRW scheme and the full GR scheme, the simulations are running with a periodic box whose size is $L=50m^{-1}$. 

\begin{figure}[htb]
  \centering
      \includegraphics[width=0.4\textwidth]{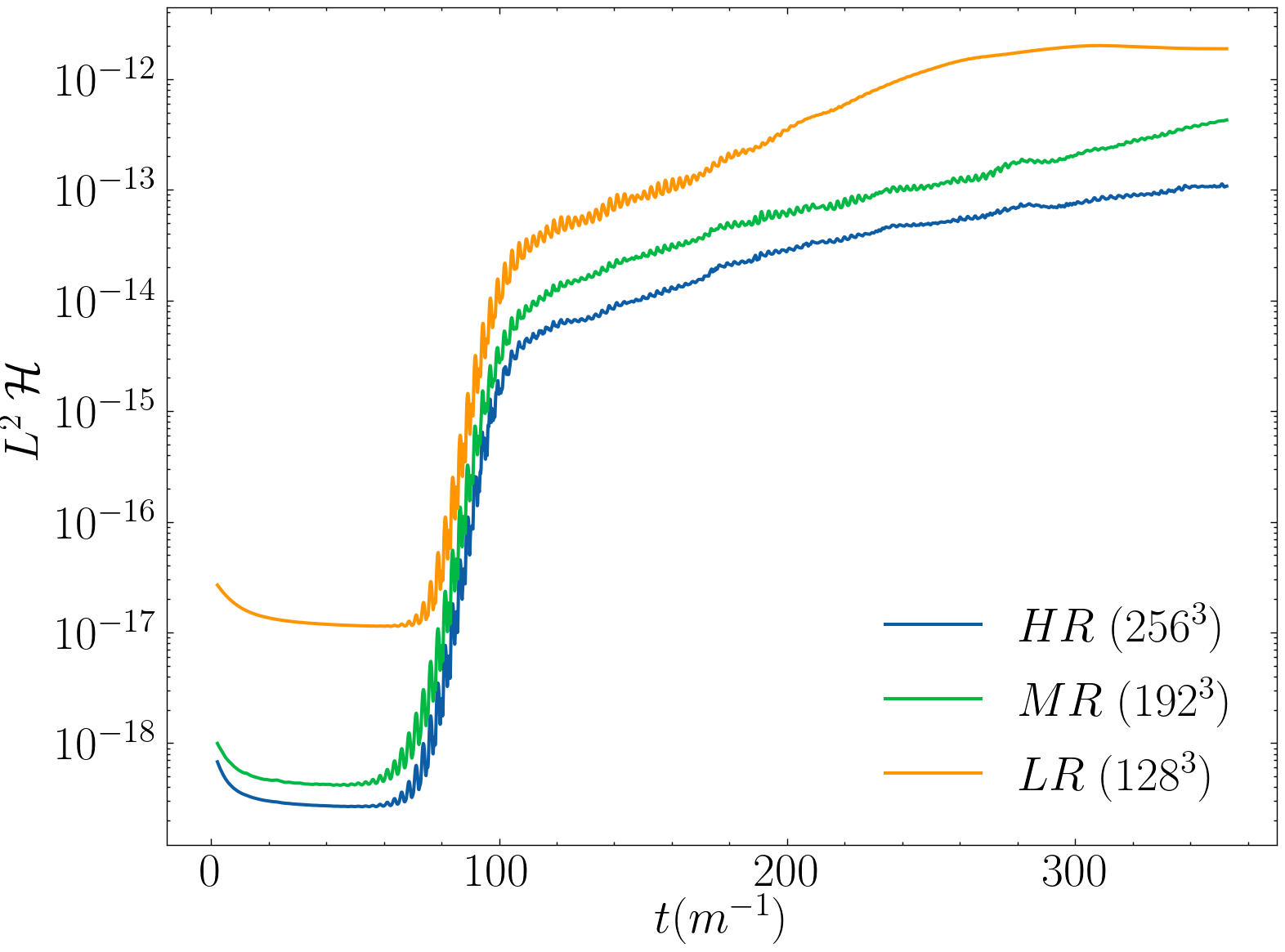}
      \caption{Behavior of the L2 norm of the Hamiltonian constraint
        under different resolutions. It corresponds the
        potential parameter combination of $\alpha=0.18$ and
        $\beta = 22$, when the GR effects are the strongest.
     A second order of convergence can be achieved.} 
    \label{H_conv}
\end{figure}

\textsc{CosmoGRaPH} employs a fourth-order Runge-Kutta scheme and a finite difference stencil with the same order. The Courant-Friedrichs-Lewy number $C$ is set to $0.2$ to maintain stability.
We will mostly use a uniform resolution of $N_{\rm res}=256$ to run our simulations. However, for cases where black holes emerge, it is crucial that the AMR feature is enabled, as we will do in Section \ref{sec:BH}. For cases where resolution is the bottleneck to achieve accurate results, we will use the linear Richardson extrapolation method to accelerate the convergence, which can be done by using extra runs with lower resolutions. That is, when extrapolations appear to be needed, we run two extra resolutions ($128^3$ and $192^3$) for the same problem, in addition to the $256^3$ run, and estimate the true value by using a convergence rate of
\begin{eqnarray}
\label{eq:c}
 c \equiv \frac{|f_{128} - f_{192}|}{|f_{192} - f_{256}|},
\end{eqnarray}
where $f_{128}$, $f_{192}$ and $f_{256}$ are the values of the quantity in question obtained from runs with resolution $128^3$, $192^3$ and $256^3$ respectively. In particular, we will need to compute $c$ for Fig.~\ref{fraction}. Typically, we find that our runs can achieve a second order convergence rate, which corresponds to $c\simeq 2.9$. For example, in Fig.~\ref{H_conv}, we show the L2 norm of the Hamiltonian constraint calculated at different resolutions, and we see that errors decrease and scale as $\Delta x^2$, indicating that the convergence rate is $c\simeq 2.9$ and thus a second order of convergence rate is achieved. All the production runs of our simulations were performed on 4 Haswell $2.67\,\rm GHz$ nodes with 24 cores per node. We make use of a hybrid of 4 OpenMPI tasks and 6 OpenMP threads on each node. The execution time for each run varies from 2,000 to 16,000 CPU hours depending on the choice of parameters.

\begin{figure}[htb]
  \centering
    \begin{subfigure}{}
      \includegraphics[width=0.215\textwidth]{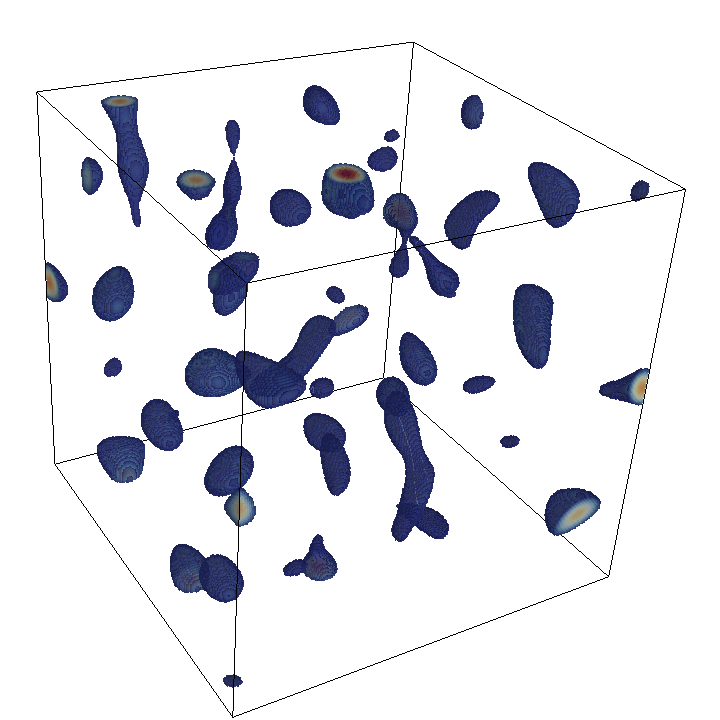}
    \end{subfigure}
    \begin{subfigure}{}
      \includegraphics[width=0.235\textwidth]{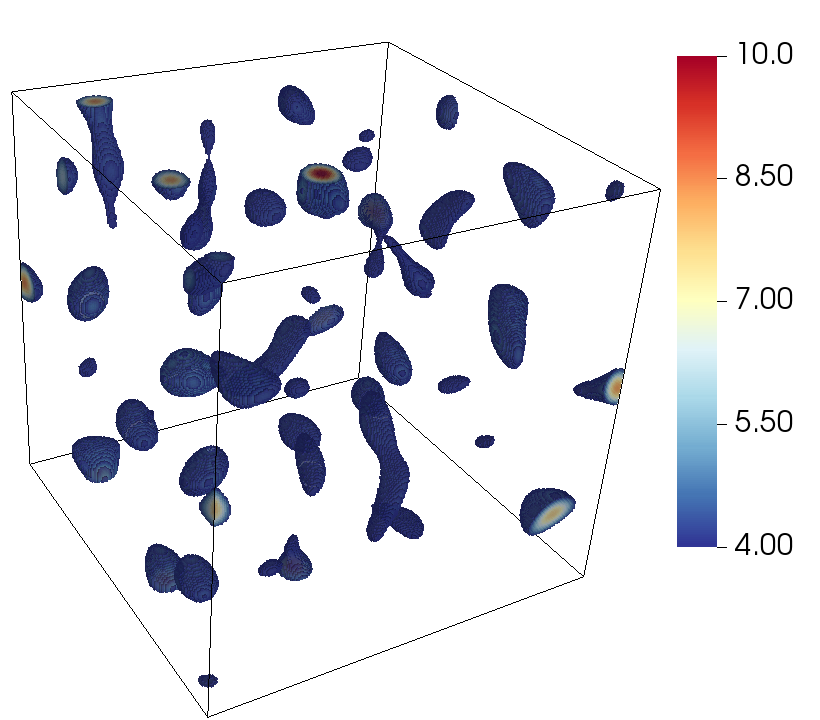}
    \end{subfigure}
    \\
  \centering
    \begin{subfigure}{}
      \includegraphics[width=0.21\textwidth]{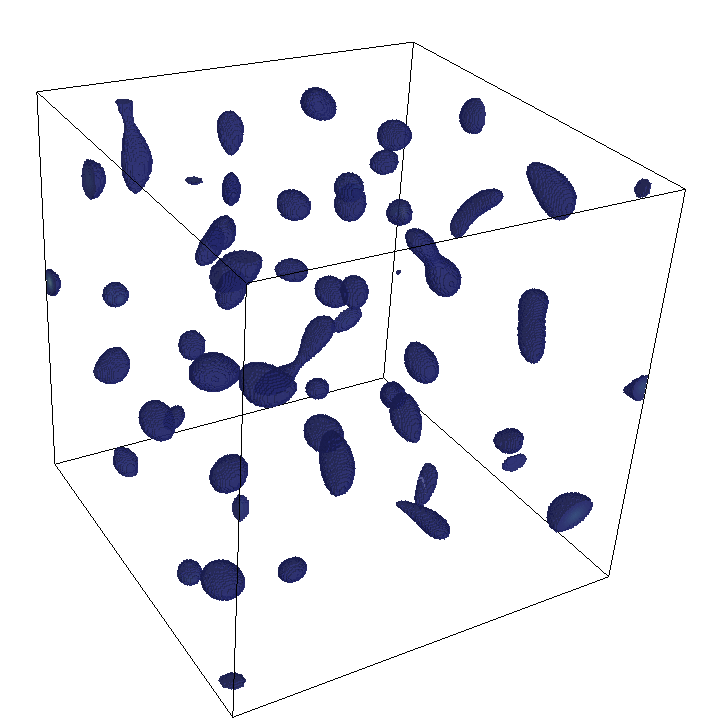}
    \end{subfigure}
    \begin{subfigure}{}
      \includegraphics[width=0.235\textwidth]{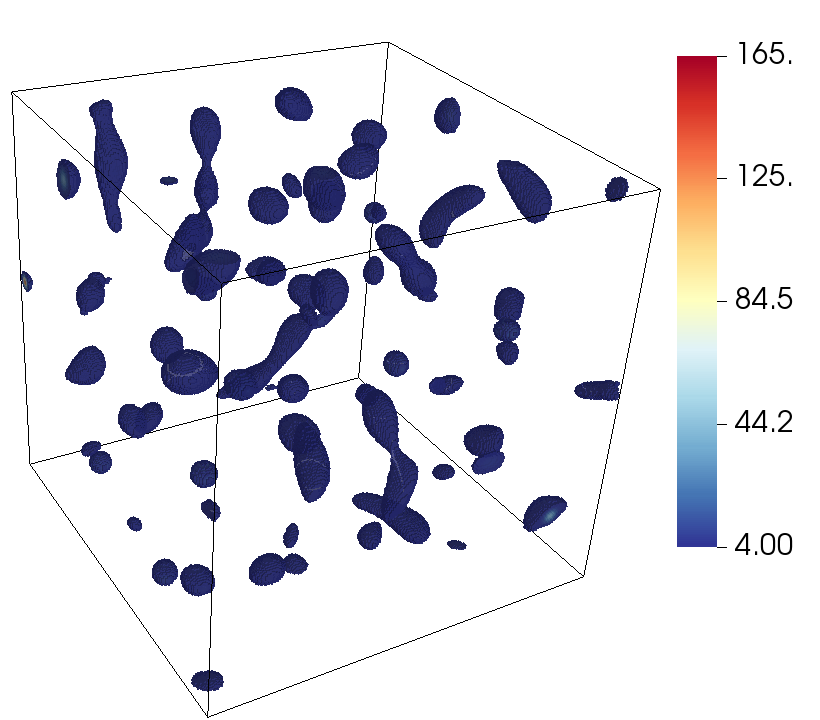}
    \end{subfigure}
    \caption{Density contrast $\rho / \left<\rho\right>$ for $\beta = 75$ (top plots)
      and $\beta=20$ (bottom plots). The left plots are simulations with the FLRW scheme,
      whereas the right plots are with the full GR scheme.
      The FLRW and GR plot with the same $\beta$ are sliced at the time when the oscillon fraction $f$ (see Eq.~(\ref{eq:2})) in the relevant full GR scheme reaches $0.28$.
      $\alpha=0.18$ for all the cases.
      } \label{rho_rho_avg}
\end{figure}

\section{Oscillon preheating in full numerical relativity}
\label{sec:1}

Now, we are ready to present the results of our simulations. In the oscillon preheating scenario, as revealed by previous simulations with the FLRW scheme, when the slow-roll parameters approach unity and the Hubble parameter drops below the mass of the inflaton, the inflaton condensate starts to oscillate, and parametric resonance takes place, fragmenting the homogeneous condensate into lumps, which then evolve to form oscillons. They are generated at almost fixed spatial positions and become stabilized after several Hubble times. They soon dominate the universe and can last for a long time, delaying thermalization. The big picture of oscillon preheating stands in the full GR simulations. Indeed, when $\beta$ is sufficiently large, the production rate of oscillons in the full GR scheme is essentially the same as in the FLRW simulations; see the top two plots of Fig.~\ref{rho_rho_avg} for the density contrast $\rho / \left<\rho\right>$ in the two schemes when $\beta = 75$ at $t=65 m^{-1}$. However, significant discrepancies manifest for smaller values of $\beta$ as the strength of the parametric resonance starts to decrease. In particular, for $\beta = 20$ at $t=345 m^{-1}$, a significantly higher production rate of oscillons can be seen when the full GR effects are taken into account, as can be visually seen in the bottom two plots of Fig.~\ref{rho_rho_avg}. This implies that strong gravity effects become important when $\beta$ is small. (Note that in Fig.~\ref{rho_rho_avg} the FLRW and GR plot with the same $\beta$ are sliced at the time when the oscillon fraction $f$ (to be more precisely defined in Eq.~(\ref{eq:2})) in the relevant full GR scheme reaches $0.28$. As parametric resonance is stronger for greater $\beta$, the $\beta = 20$ GR scheme at $t=345 m^{-1}$ reaches the same oscillon fraction as the $\beta = 75 $ GR case at $t=65 m^{-1}$, so the energy density in the left bottom plot is visually less condensed than that in the left top plot.)

\begin{figure}[htb]
  \centering
      \includegraphics[width=0.4\textwidth]{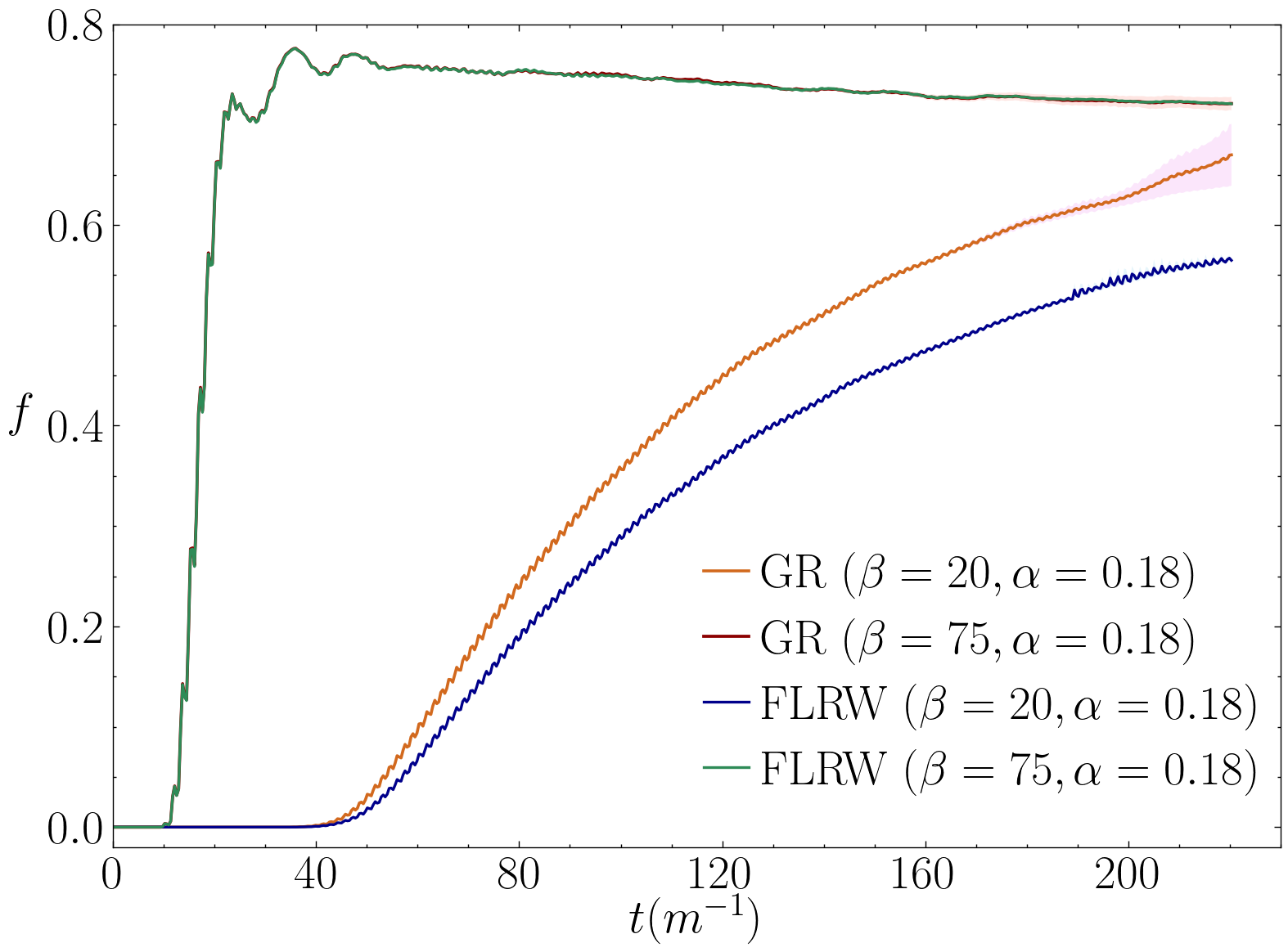}
      \caption{Time evolution of the fractions of energy contained in oscillons. ``GR'' stands for the full GR scheme and ``FLRW'' for the FLRW scheme. Error bars are denoted using shaded bands, and are not visible when very small. The $\beta=75$ GR and FLRW lines mostly overlap.}
    \label{fraction_evolve}
\end{figure}

This difference can be quantified by computing the energy fraction the oscillons contain, which is defined as the fraction of energies contained in regions where the energy density is greater than twice the average:
\begin{align}
\label{eq:2}
f=\frac{\int_{\rho > 2 \left<\rho\right>} \rho \mathrm dV}{\int \rho \mathrm dV}.
\end{align}
In the sense of general relativity, the energy density is of course a frame dependent quantity. However, as discussed in the last section, we have chosen the initial and gauge conditions such that initially the frame mostly coincides with that of the FLRW case, and in the consequent evolution neither the spacetime as a whole drifts away from a FLRW background nor do the gauge conditions significantly deviate from the initial ones at large scales. In other words, physically, there is still an FLRW background, so it is meaningful to compare energy densities with the FLRW scheme.

As can be seen in Fig.~\ref{fraction_evolve}, when $\beta$ is large and parametric resonance is very strong, the fraction increases quickly with time until reaching a plateau around $f=0.7$, with the fraction in the full GR scheme agreeing very well with the FLRW scheme. On the other hand, discrepancies emerge for smaller $\beta$ such as $\beta=20$, in which case parametric resonance is relatively weaker and the fraction grows slower with time, but the effects of self-gravity become much more important and a significantly higher fraction of oscillons is observed in the full GR scheme. The shaded region represents uncertainties from the extrapolations from runs with different resolutions (see \cite{oberkampf2010verification} for the extrapolation and error estimation method).

The energy fraction $f$ is defined in Eq.~\eqref{eq:2} with the threshold $2\langle \rho \rangle$. To be prudent, we also compare the oscillon fractions with thresholds higher than $2$, and we get the same conclusion. For example, if setting the threshold to $5$, we have: when $\beta = 75$, the oscillon fractions reach $0.691\pm0.0036$ and $0.6903\pm0.0041$ for the FRW and GR scheme respectively; while for a smaller $\beta = 22$, the difference is significant again, the fraction for the FRW scheme being $0.618\pm0.002$, compared to $0.687\pm0.042$ in the GR scheme.

\begin{figure}[htb]
 \centering
      \begin{subfigure}{}
      \includegraphics[width=0.4\textwidth]{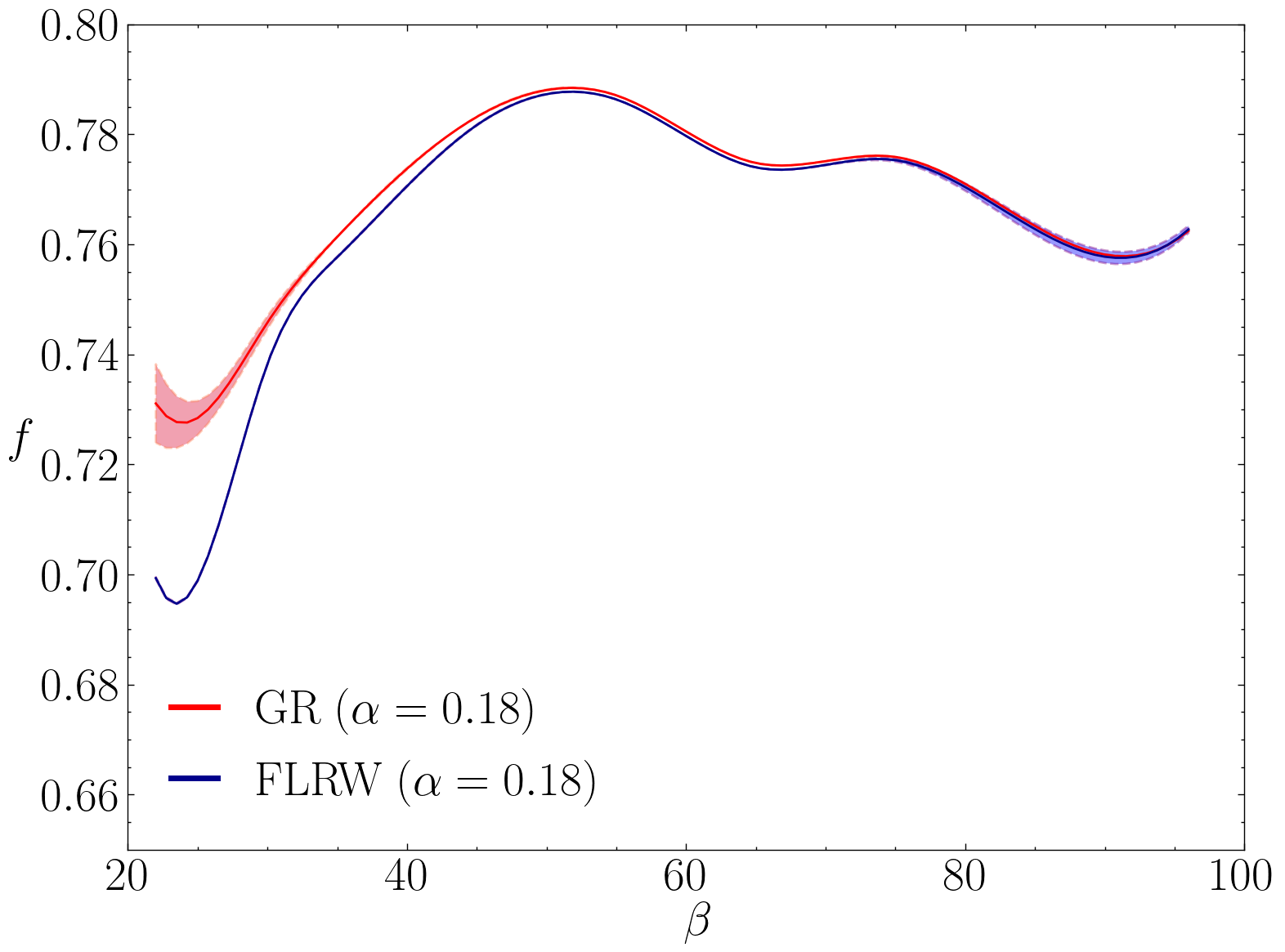}
    \end{subfigure}
    \begin{subfigure}{}
      \includegraphics[width=0.4\textwidth]{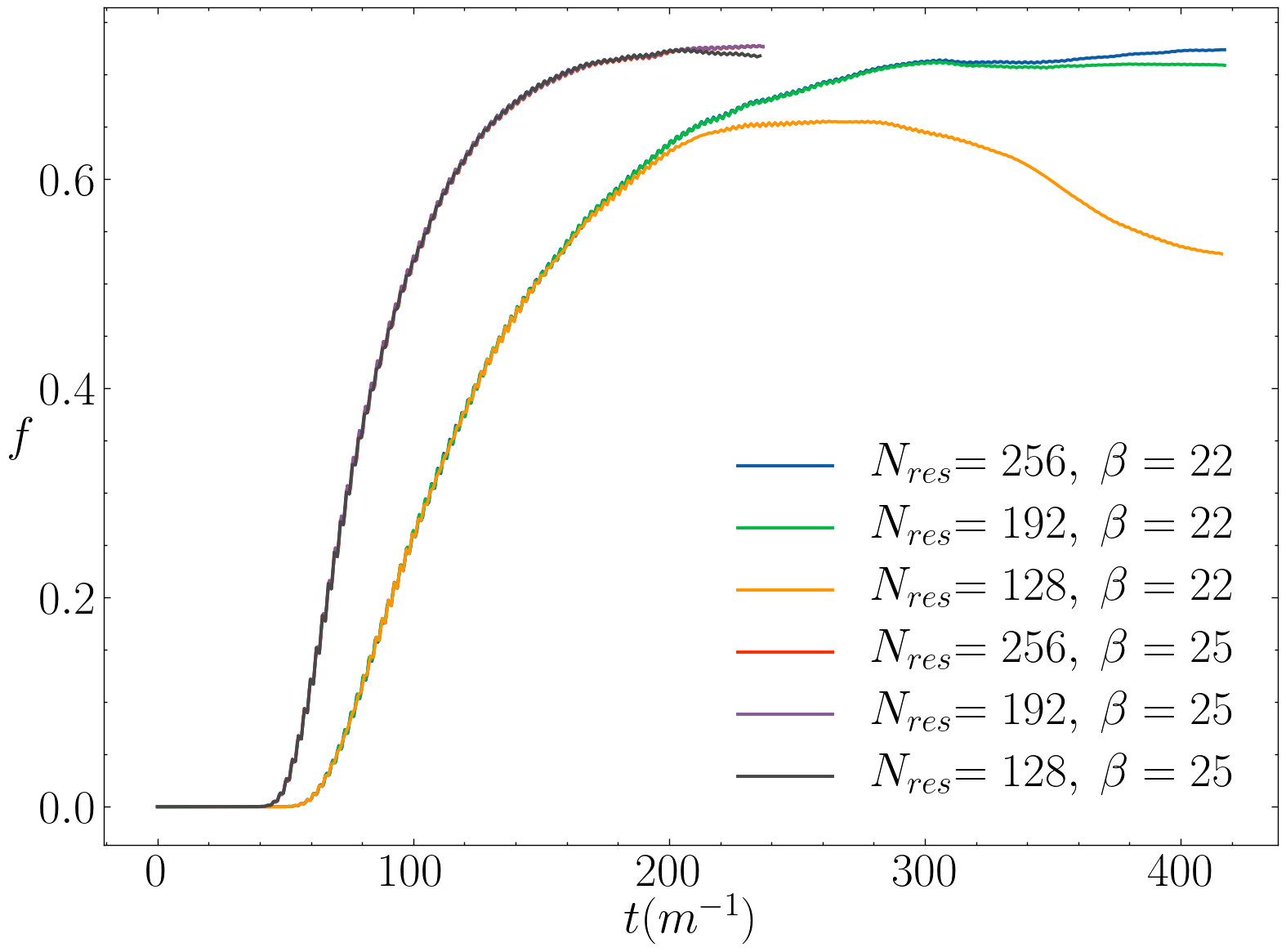}
    \end{subfigure}
      \caption{Comparisons of the maximum fractions of energy that oscillons contain for $\beta=22$ to $95$ in the FLRW and full GR scheme. The convergence tests for the two full GR simulations that have most numerical instabilities ($\beta=22$ and $\beta=25$) are showed in the right plot.}
    \label{fraction}
\end{figure}

Also, see Fig.~\ref{fraction} for a comparison of the maximum oscillon fractions in the
two schemes for different $\beta$'s. For a given $\beta$, the central value for either scheme is obtained by extrapolating the value of the maximum fraction among different resolutions during the time-evolution, and the errors represented by the shadow region are from these the extrapolating processes. Specifically, the error shadows are obtained from the discrepancies between the extrapolated values (with convergence rate $c$ as defined in Eq.~(\ref{eq:c})) and the high resolution values. The fraction of energy in the FLRW scheme agrees very well with the previous work \cite{Amin:2011hj} and coincides perfectly with the full GR scheme when $\beta$ is large. However, although with increasing uncertainties, the central fraction of energy in the full GR scheme gradually deviates from
that of the FLRW scheme when $\beta$ decreases. Thus, again, we see that a fully general-relativistic treatment becomes essential for small $\beta$.

The increasing trend of the extrapolating errors in the full GR scheme can also be seen from the convergence tests for two of the most unstable GR simulations ($\beta=22$ and $\beta=25$, as shown in the right plot in Fig.~\ref{fraction}): although still exhibiting a second order convergence, the $\beta=22$ case converges slower than the $\beta=25$ case when the fraction reaches its maximum. Below $\beta = 22$, the GR code running with the two high resolutions could stably resolve the dynamics and greater deviations from the FLRW scheme can be seen when $\beta$ decreases, but the convergence rate falls below second order before the maximum of the oscillon fraction is reached. The increasing difficulty to resolve the dynamics with a decreasing $\beta$ suggests that self-gravity is so strong in the small $\beta$ regime that the resolution is becoming the bottle neck to improve the accuracy of the simulation. In the next section, we will make use of the AMR functionality to resolve this difficulty for small $\beta$ when the resolution is insufficient.

\begin{figure}[htb]
  \centering
      \begin{subfigure}{}
      \includegraphics[width=0.4\textwidth]{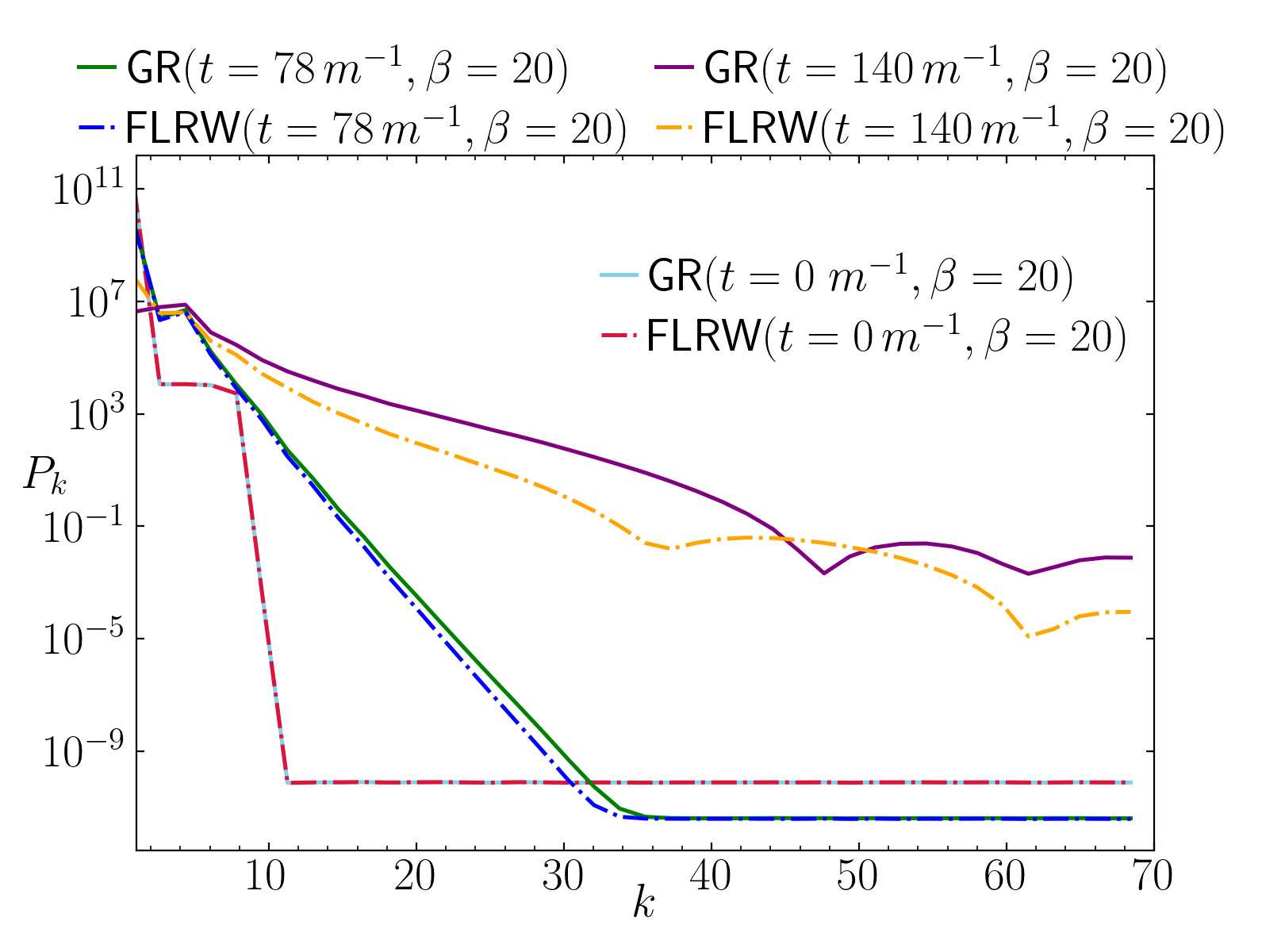}
    \end{subfigure}
    \begin{subfigure}{}
      \includegraphics[width=0.4\textwidth]{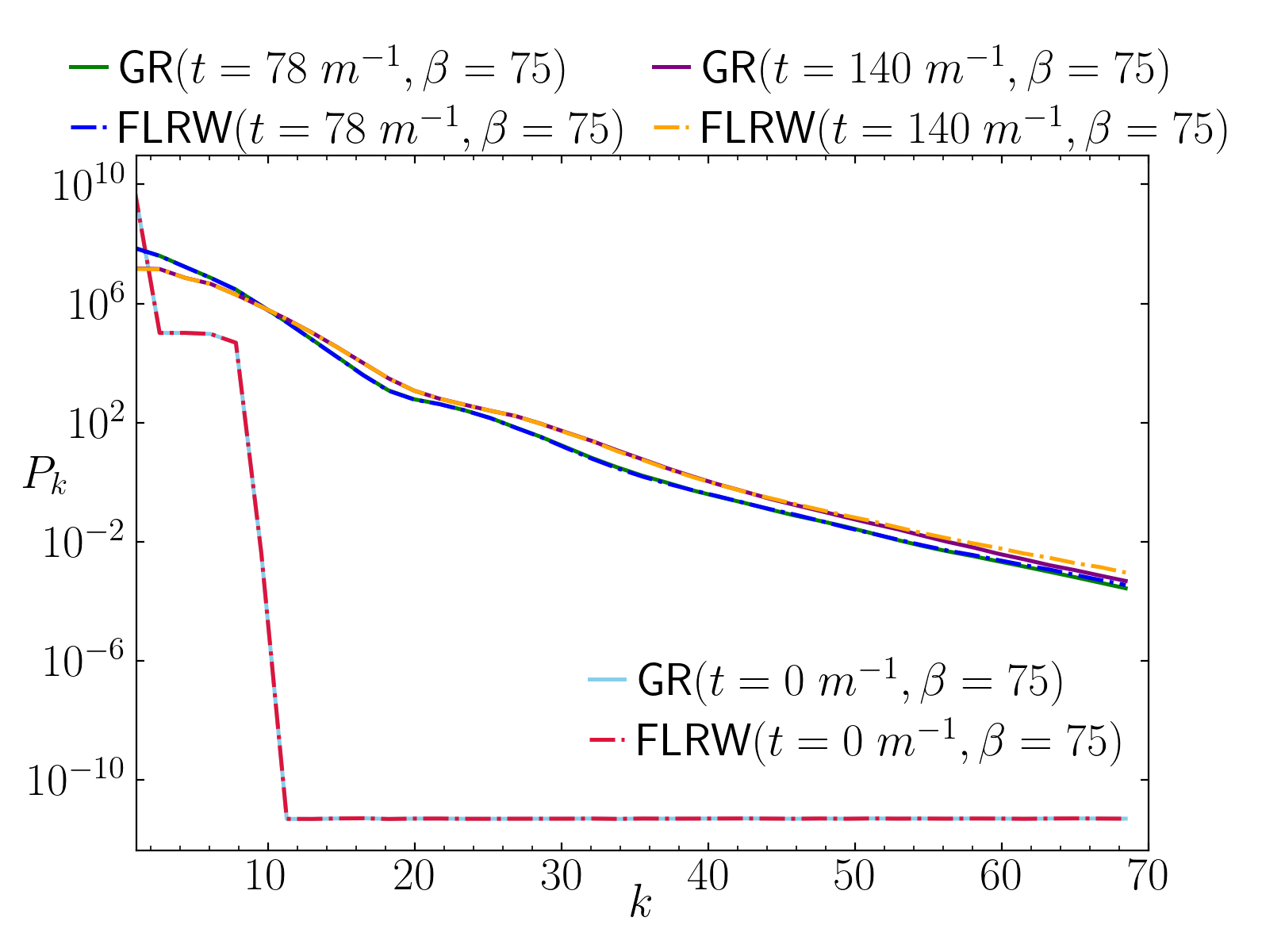}
    \end{subfigure}
    \caption{Comparisons of the power spectrum of $\phi$ at different time slices between the FLRW and full GR scheme with different $\beta$.
     $\alpha$ is chosen to be $0.18$.}
       \label{power_diff}
\end{figure}

The strong self-gravity effects for small $\beta$ can also effectively back-react to the scalar sector.
To confirm this, we calculate the power spectrum $P_k$ of $\phi$.  As shown in Fig.~\ref{power_diff}, the power spectrum of the scalar field is significantly higher in the full GR scheme than in the FLRW scheme, whereas for large $\beta$ the two schemes are almost the same except for high $k$.

\section{Refinements with AMR}
\label{sec:amr}

\begin{figure}[h]
 \centering
      \begin{subfigure}{}
      \includegraphics[width=0.21\textwidth]{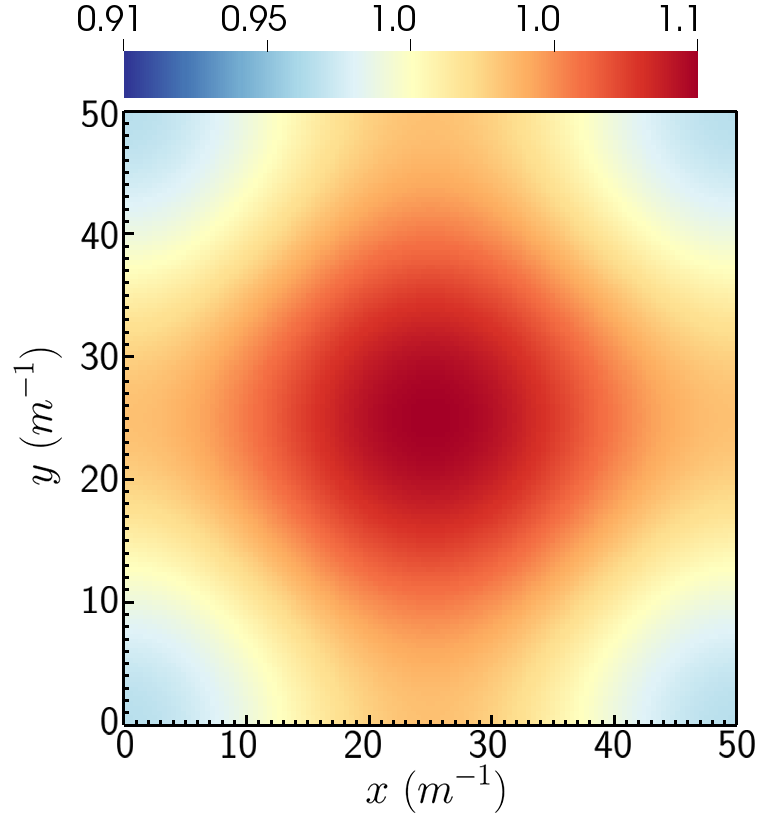}
    \end{subfigure}
    \begin{subfigure}{}
      \includegraphics[width=0.21\textwidth]{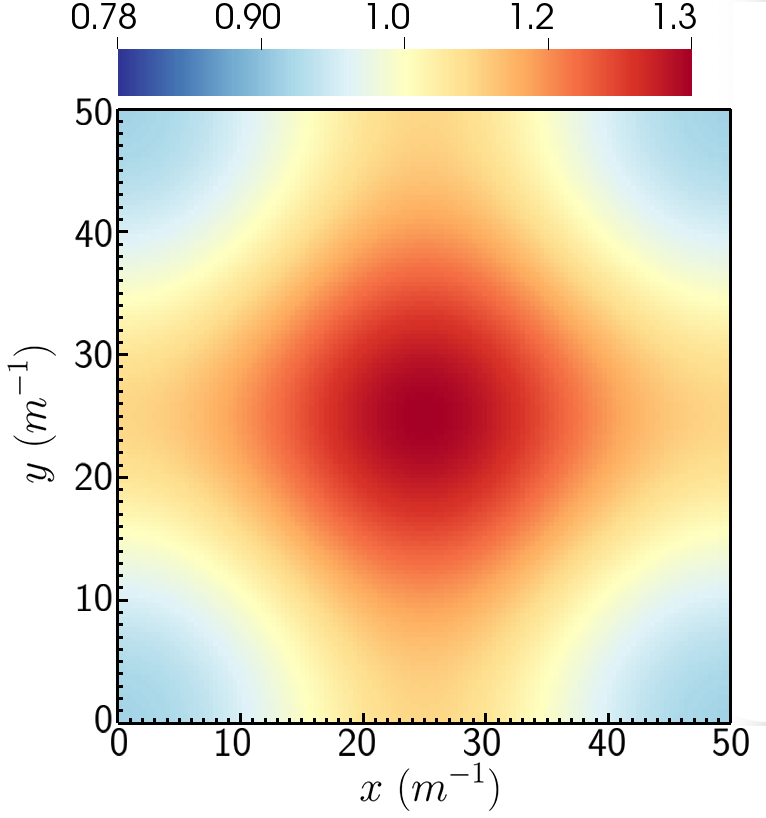}
    \end{subfigure}
    \begin{subfigure}{}
      \includegraphics[width=0.21\textwidth]{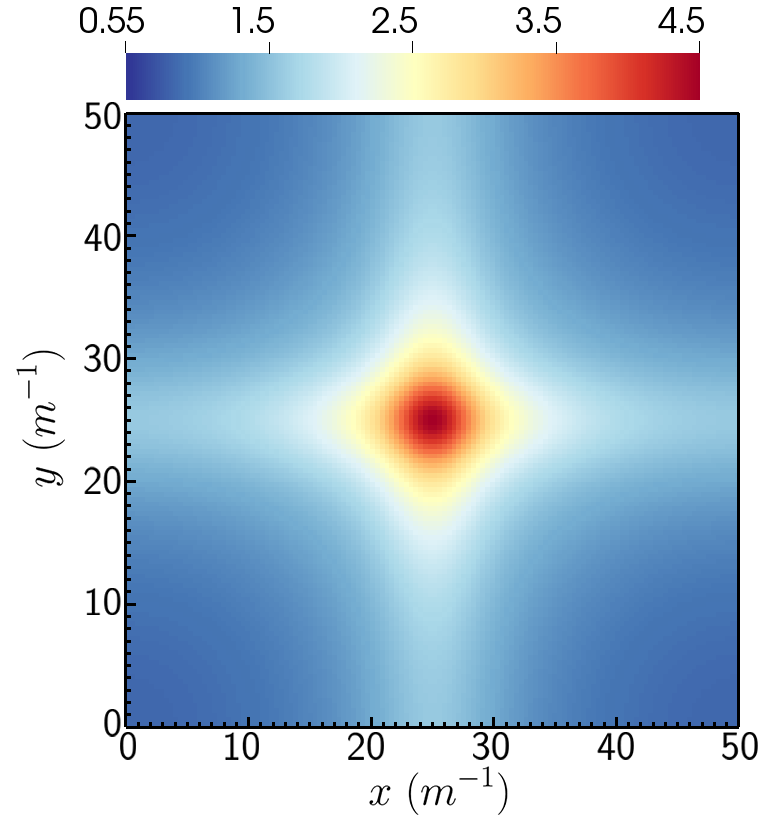}
    \end{subfigure}
    \begin{subfigure}{}
      \includegraphics[width=0.21\textwidth]{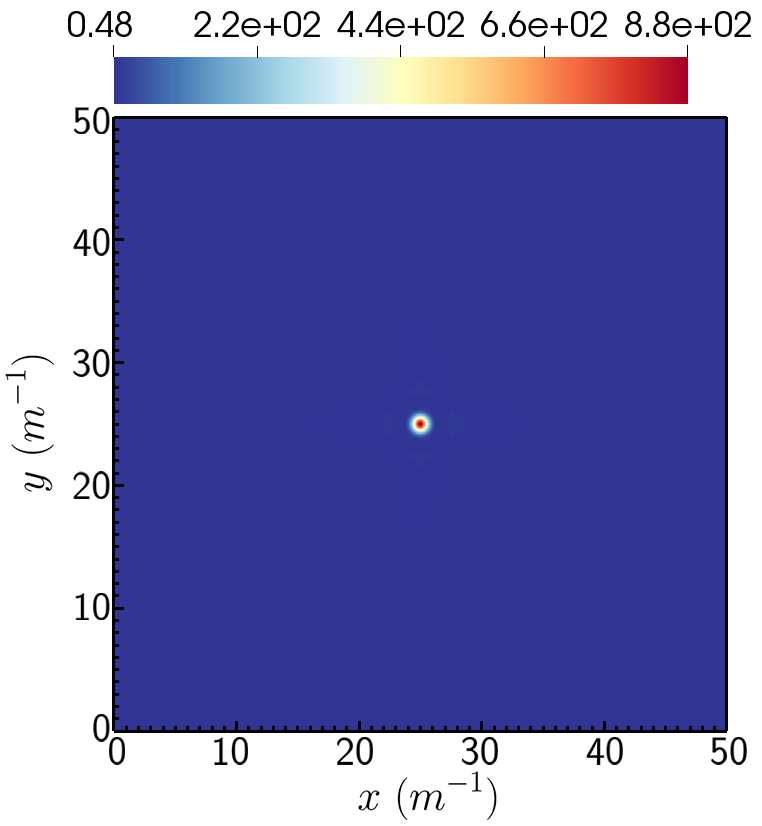}
    \end{subfigure}

    \begin{subfigure}{}
      \includegraphics[width=0.21\textwidth]{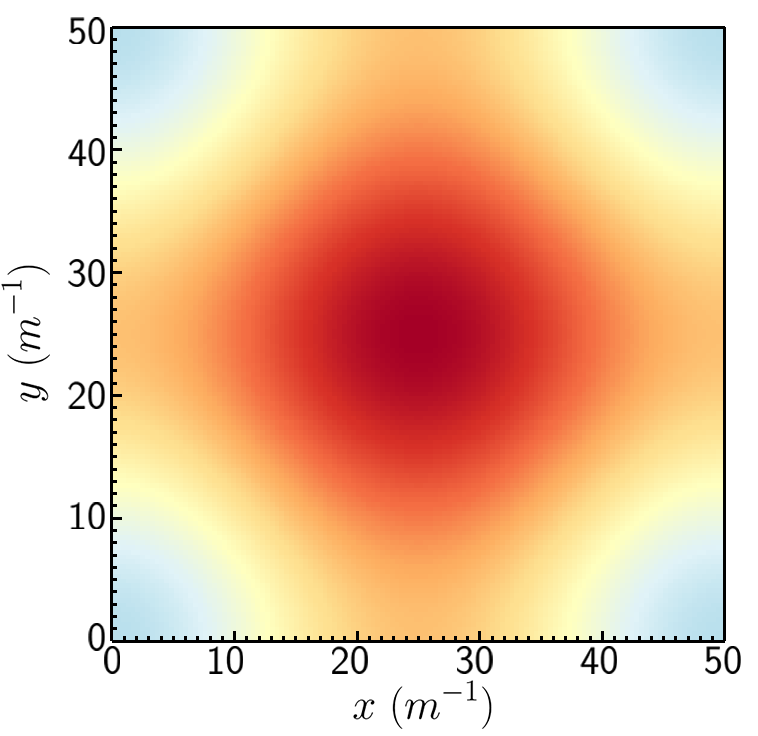}
    \end{subfigure}
    \begin{subfigure}{}
      \includegraphics[width=0.21\textwidth]{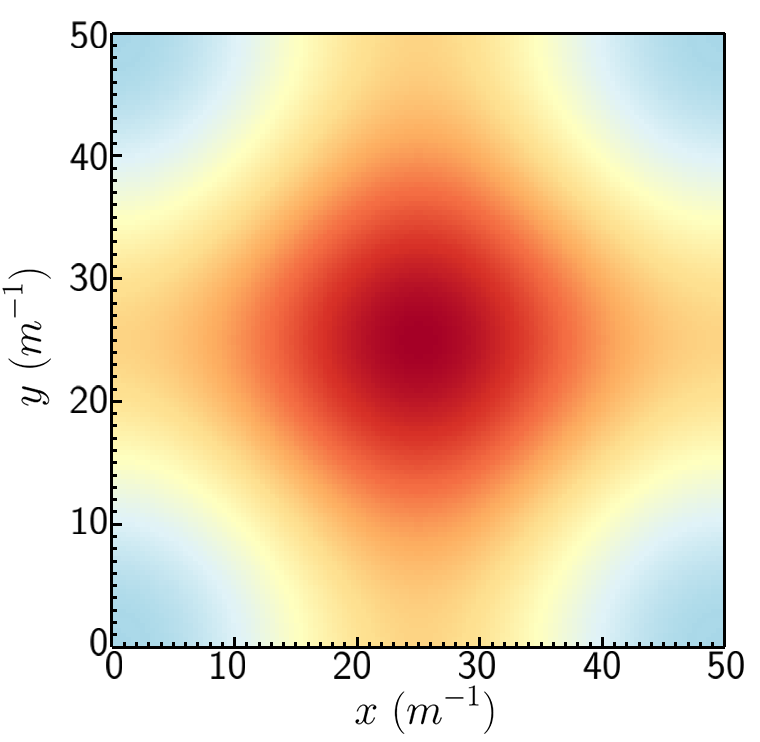}
    \end{subfigure}
    \begin{subfigure}{}
      \includegraphics[width=0.21\textwidth]{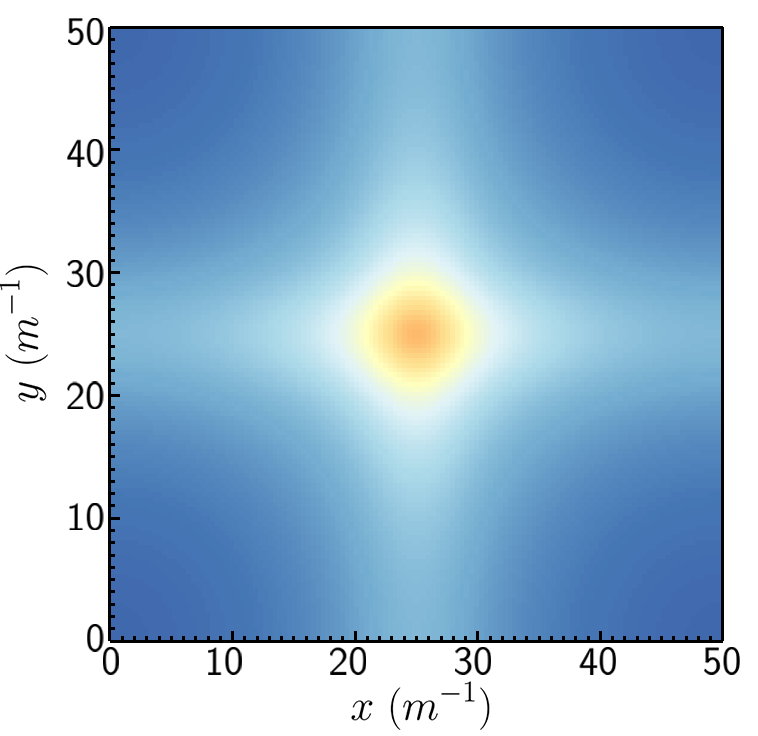}
    \end{subfigure}
    \begin{subfigure}{}
      \includegraphics[width=0.21\textwidth]{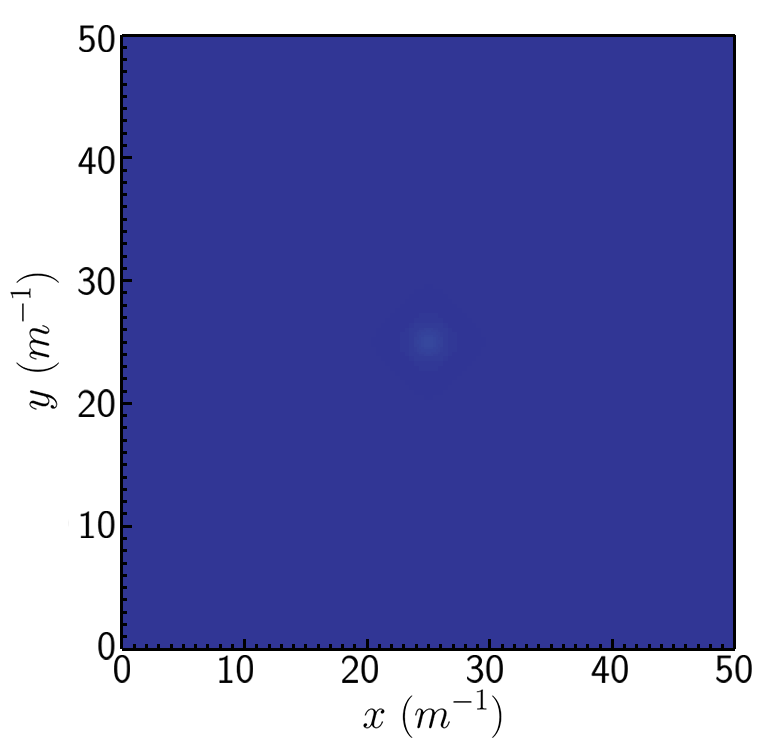}
    \end{subfigure}

     \begin{subfigure}{}
      \includegraphics[width=0.21\textwidth]{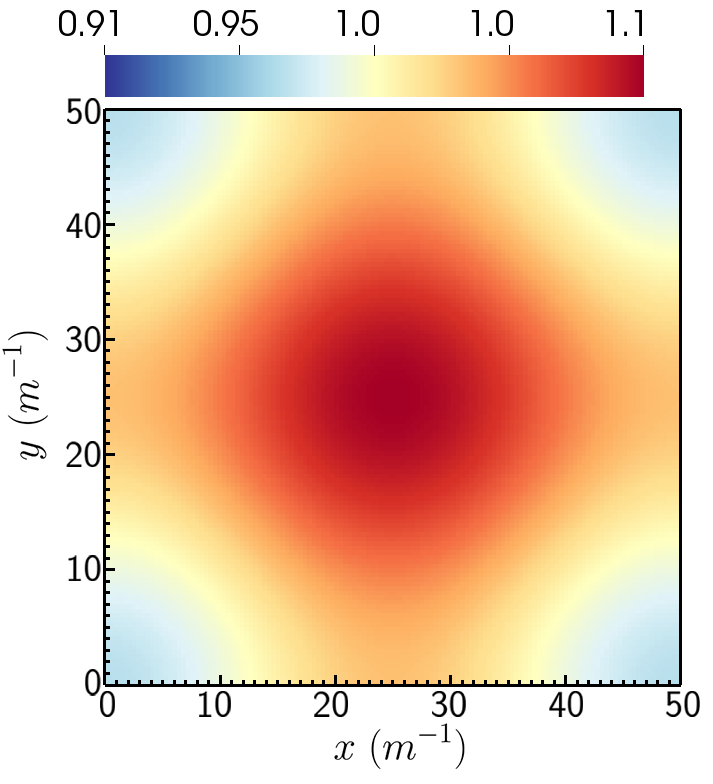}
    \end{subfigure}
    \begin{subfigure}{}
      \includegraphics[width=0.21\textwidth]{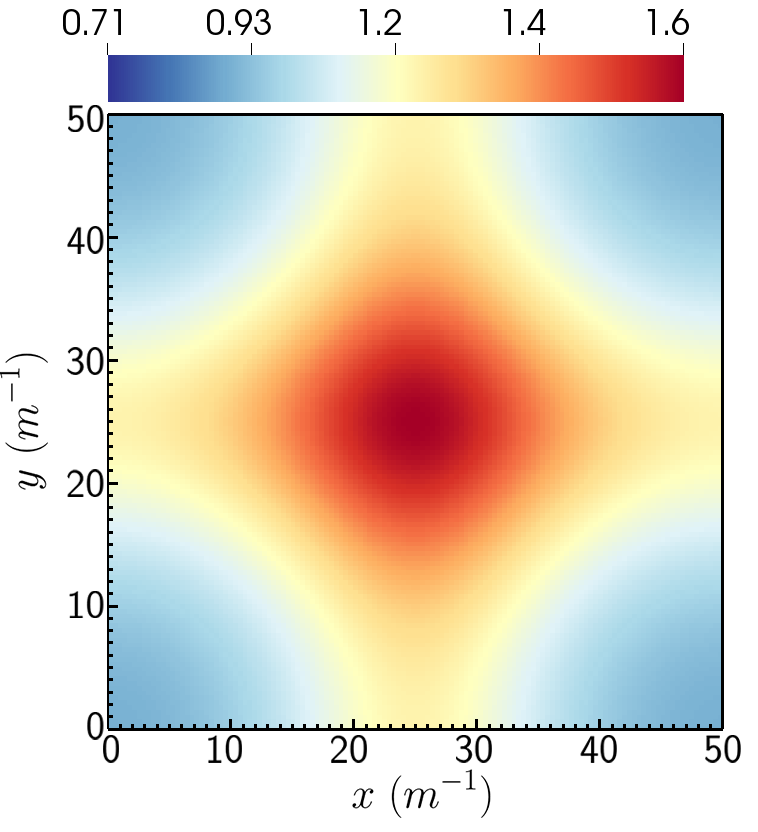}
    \end{subfigure}
    \begin{subfigure}{}
      \includegraphics[width=0.21\textwidth]{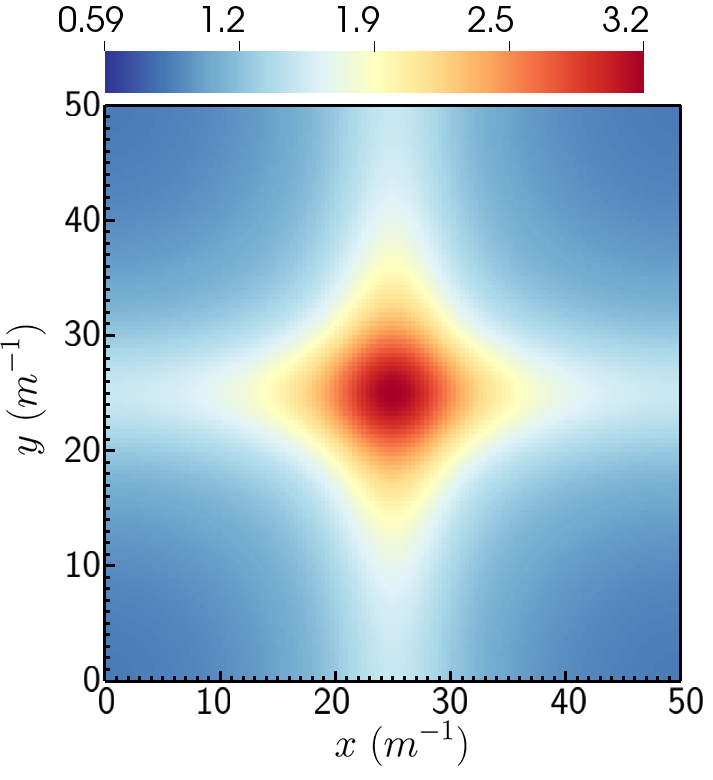}
    \end{subfigure}
    \begin{subfigure}{}
      \includegraphics[width=0.21\textwidth]{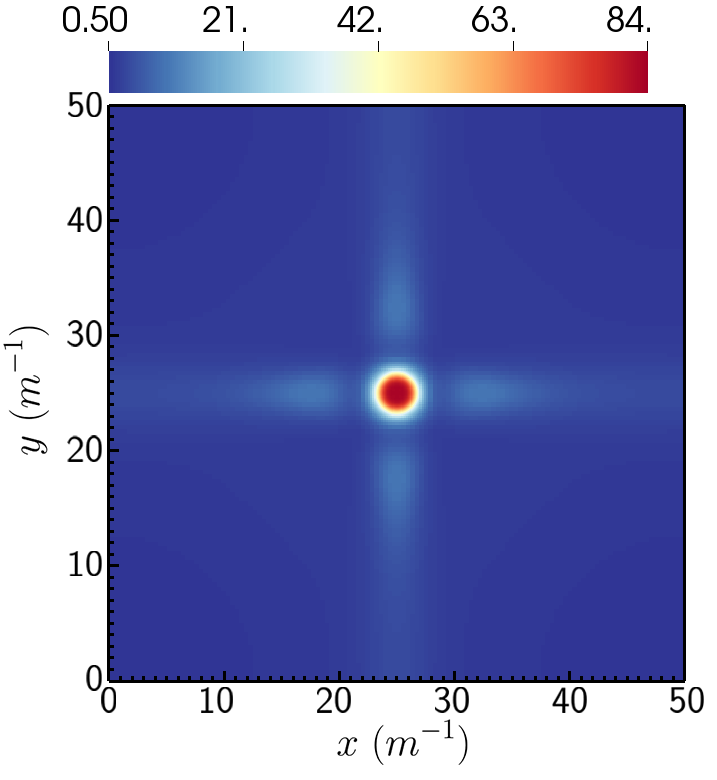}
    \end{subfigure}

    \begin{subfigure}{}
      \includegraphics[width=0.21\textwidth]{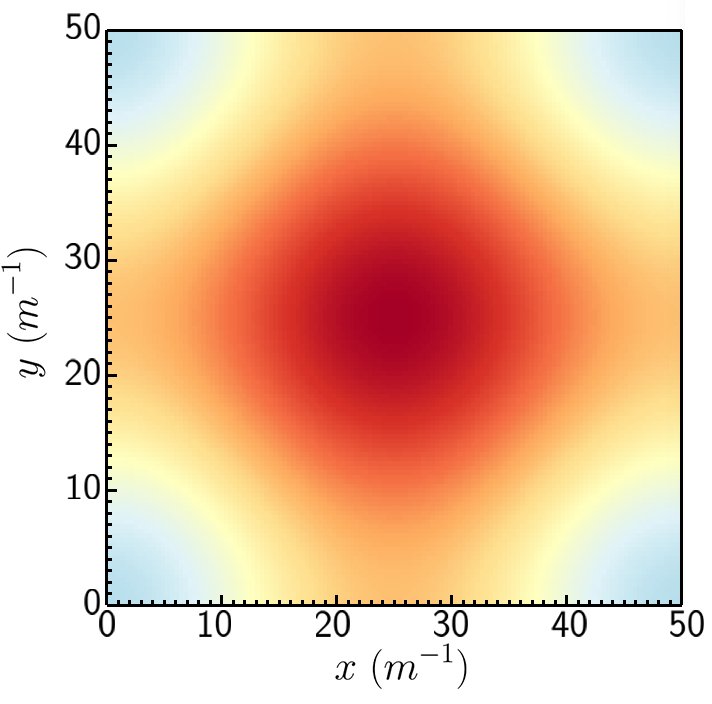}
    \end{subfigure}
    \begin{subfigure}{}
      \includegraphics[width=0.21\textwidth]{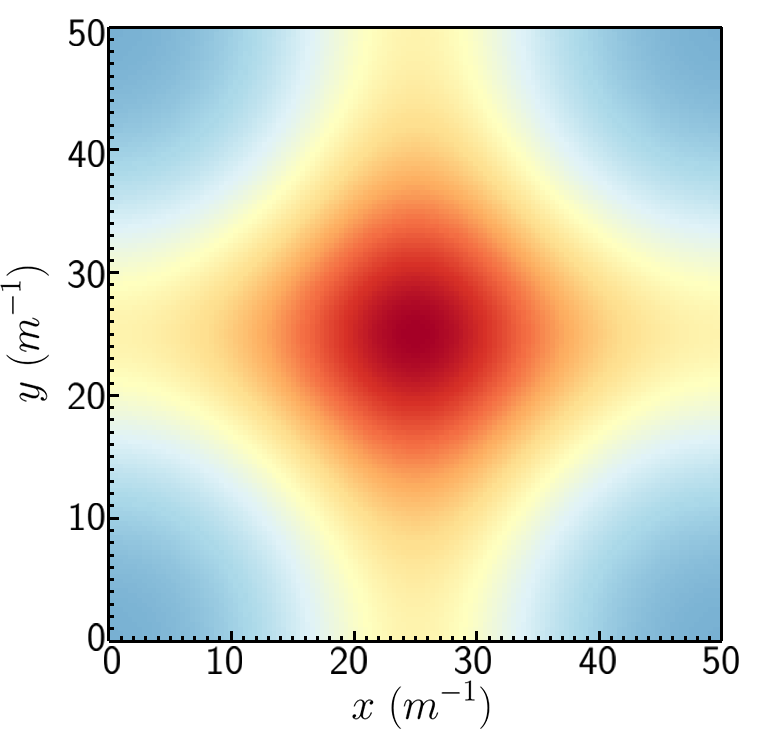}
    \end{subfigure}
    \begin{subfigure}{}
      \includegraphics[width=0.21\textwidth]{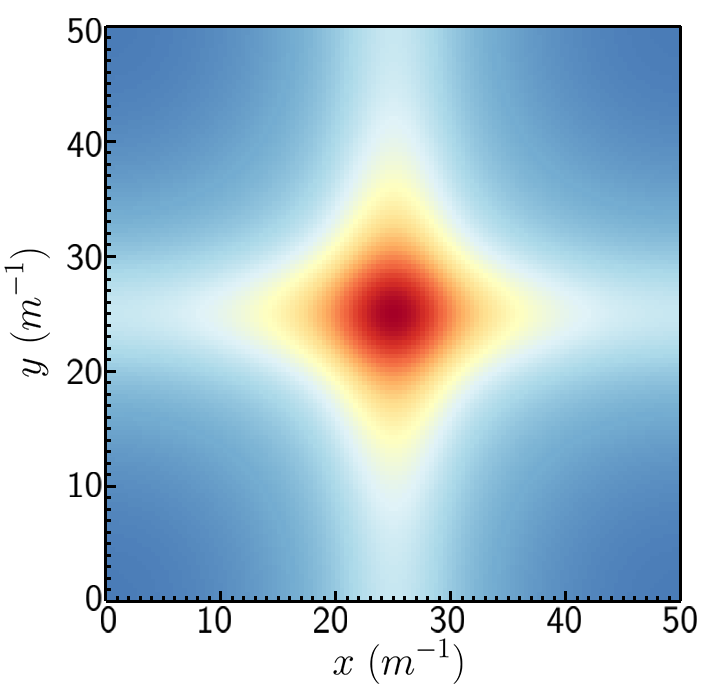}
    \end{subfigure}
    \begin{subfigure}{}
      \includegraphics[width=0.21\textwidth]{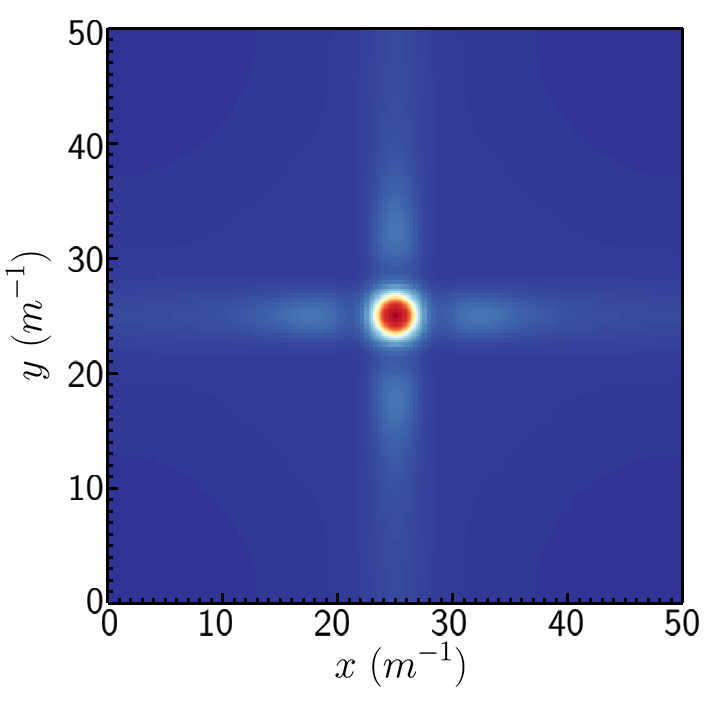}
    \end{subfigure}

    \caption{Comparisons of the time evolution of the
      density profile $\rho / \left<\rho\right>$ of a single oscillon
      between the GR and FRW scheme.
      The 1st and 2nd row correspond to the $\rho / \left<\rho\right>$ between
      the GR and the FLRW scheme respectively for $\alpha = 0.18$ and $\beta = 25$,
      and the time slices are from $t= 0$ (left) to the time when the oscillon forms
      (right).
      The 3rd and 4th row are the GR and the FLRW scheme respectively for $\alpha = 0.18$ and $\beta = 75$.
      The resolution of the bottom layer of the AMR grid is $N_{\rm res}=128$ and
      five layers of AMR levels are enabled.}
    \label{single_oscillon}
\end{figure}

By simulating oscillon generation under uniform grids, we have shown in the previous section that, for small $\bi$, nonlinear self-gravity becomes important and the energy densities of the generated oscillons are significantly increased. 
Moreover, the fact that more numerical resolutions are needed for small $\bi$ might be taken as a hint of some unresolved strong-gravity phenomena there.
It is therefore necessary to conduct some refined simulations for small $\beta$.
To this end, we turn on the AMR functionality of our case to accurately resolve the small scale non-perturbative dynamics. We use the density contrast $\rho / \left< \rho \right>$ as the indicator to mark when and where to increase the resolution so that the regions where oscillons form will have higher resolutions.
For an easy identification of the oscillon, we apply a single box-size mode as the initial data for the scalar field on the initial slice such that only a single oscillon would form in the box during the preheating process. In other words, we perform ``zoom-in'' simulations of oscillon preheating in the AMR grid.

The comparison between the evolution
of the density profile $\rho / \left<\rho\right>$ with the GR and the FLRW scheme
is presented in Fig.~\ref{single_oscillon},
where $\alpha = 0.18$, $\beta = 25$ for the top two rows and $\beta = 75 $ for the bottom two rows. It can be seen from the top two rows that for small $\beta$ the GR scheme tends to condensate more energy in the oscillon than the FRW scheme,
and faster, which further validates with the results presented in Sec. \ref{sec:1}, while for large $\beta$ the two schemes give mostly the same results.
For small $\beta$, the fact that self-gravity effects are strong and important for the formation of these oscillons suggests that the internal structure of these oscillons might be significantly different from those of large $\beta$, and they may be prone to gravitational collapses, as we will show in the following.

\subsection{Black hole formation}

\label{sec:BH}

\begin{figure}[htb]
 \centering
      \begin{subfigure}{}
      \includegraphics[width=0.4\textwidth]{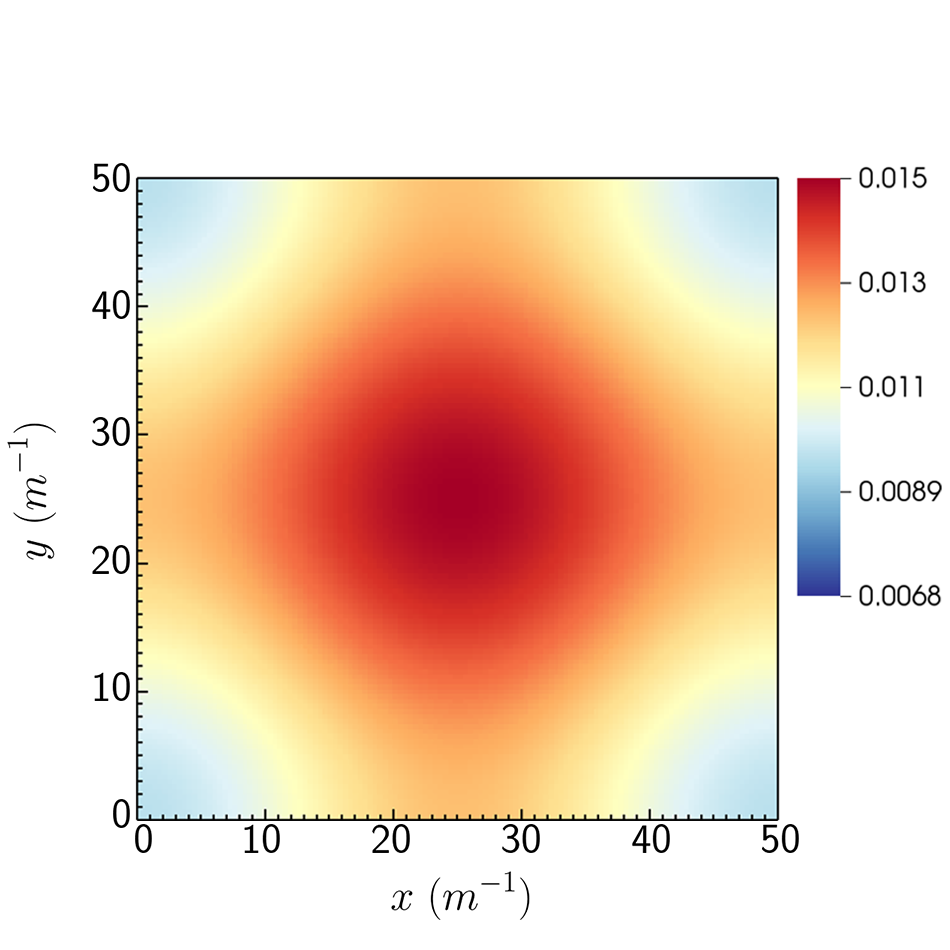}
    \end{subfigure}
    \begin{subfigure}{}
      \includegraphics[width=0.4\textwidth]{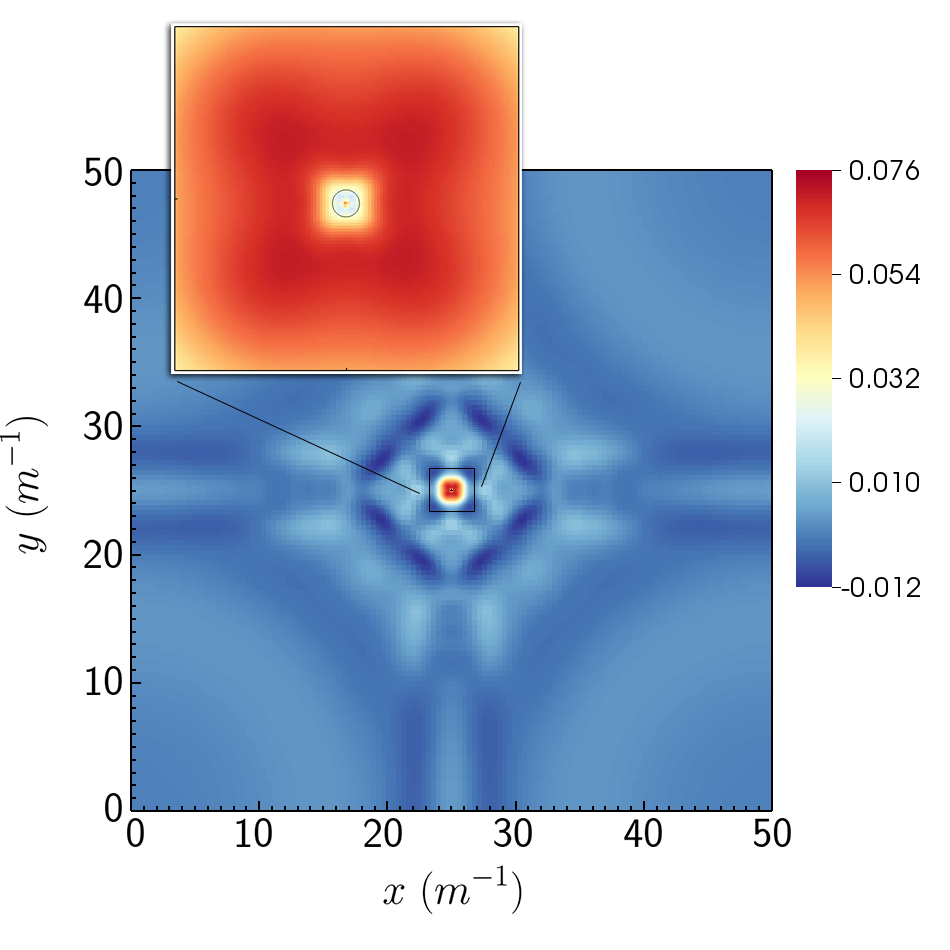}
    \end{subfigure}
    \caption{The initial slice of the scalar field $\phi$ (left plot) and 
      the slice of the scalar field when a black hole forms (right plot). The location of the apparent horizon is showed in the zoom-in subplot with a black circle. The resolution of the bottom layer is $N_{\rm res}=128$ and six layers of AMR levels are enabled.}
    \label{bh}
\end{figure}

To investigate whether the gravitational effects
can collapse the oscillons into black holes, we continue to make use of a single box-size mode for easy identification of the apparent horizon of the black hole. 
We tune the amplitude of the initial fluctuation such that the oscillon fraction before the black hole formation coincides with the oscillon fraction from an initial spectrum with more modes. Indeed, black holes can form for small $\beta$. See Fig.~\ref{bh} for an example of an apparent horizon with $\alpha = 0.18$ and $\bi = 18$: the initial slice of the scalar field (left plot) and the slice of the scalar field when an apparent horizon first appeared (right plot) are shown. 

\begin{figure}[htb]
  \centering
      \includegraphics[width=0.4\textwidth]{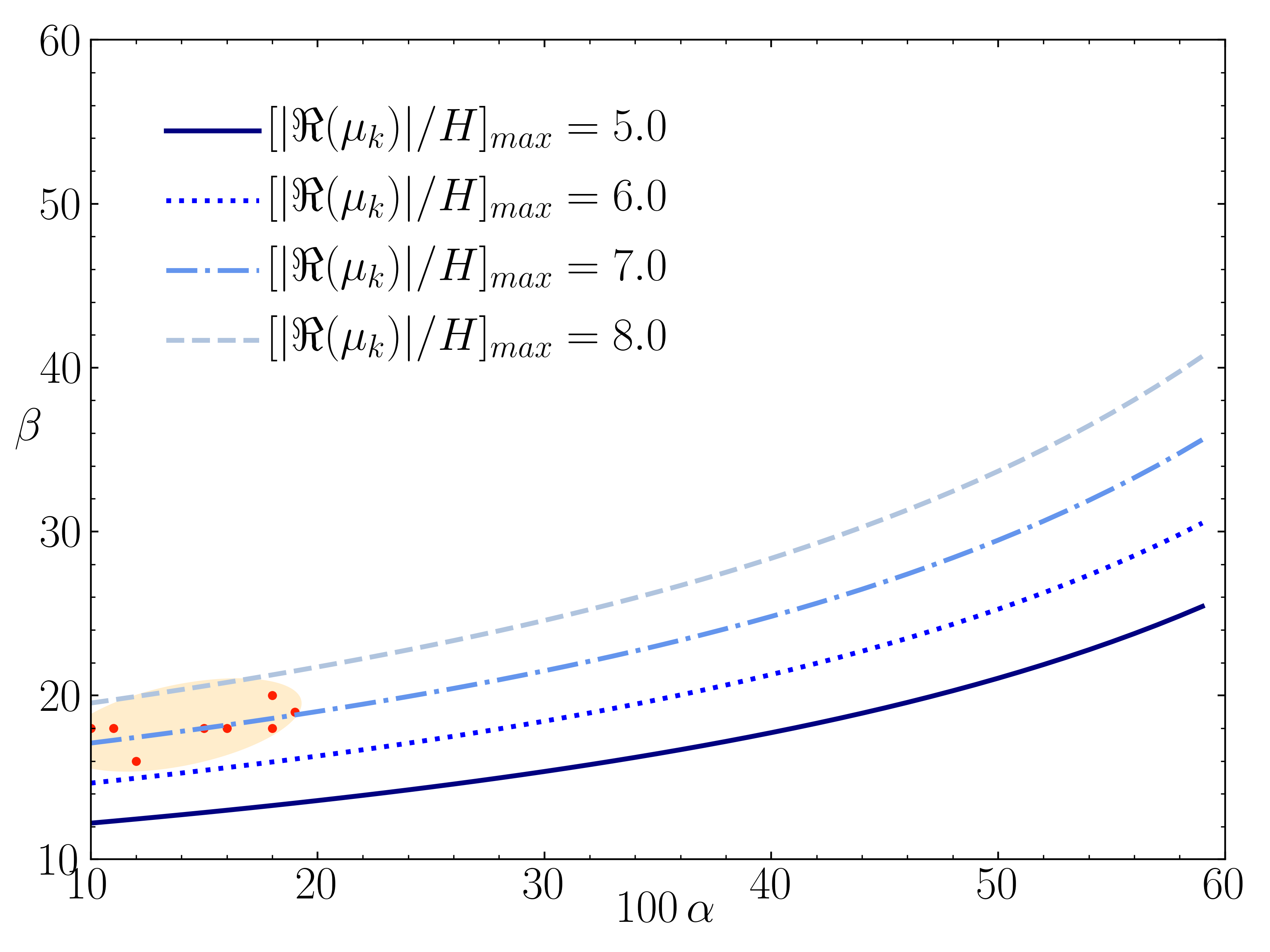}
      \caption{Tentative boundary of the parameter space where oscillons can collapse to black holes. 
Red dots are randomly chosen runs where apparent horizons can be identified for the corresponding $\alpha$ and $\beta$.}
    \label{region}
\end{figure}

We have also initiated the task to carve out the 2D parameter space ($\ai$ and $\bi$) where oscillons can collapse to black holes, which is difficult because, even with AMR, high resolutions and better convergence of the apparent horizon finder turn out to be crucial for this task. This limits our ability to mark a clear boundary for the black hole formation region. In Fig.~\ref{region}, based on about 20 runs, we have marked a tentative boundary of the black hole formation region. Since the line with $\[|\Re(\mu_k)| / H \]_{max} = 7$ approximately marks the strong resonance region in the FLRW scheme, the overlapping between the black hole region and the strong resonance region suggests extra care needs to be taken and a fully general relativistic simulation with a sufficiently high resolution will be needed in this parameter region.

\subsection{Larger black holes}
\label{sec:largeBH}

\begin{figure}[htb]
    \centering
        \includegraphics[width=0.45\textwidth]{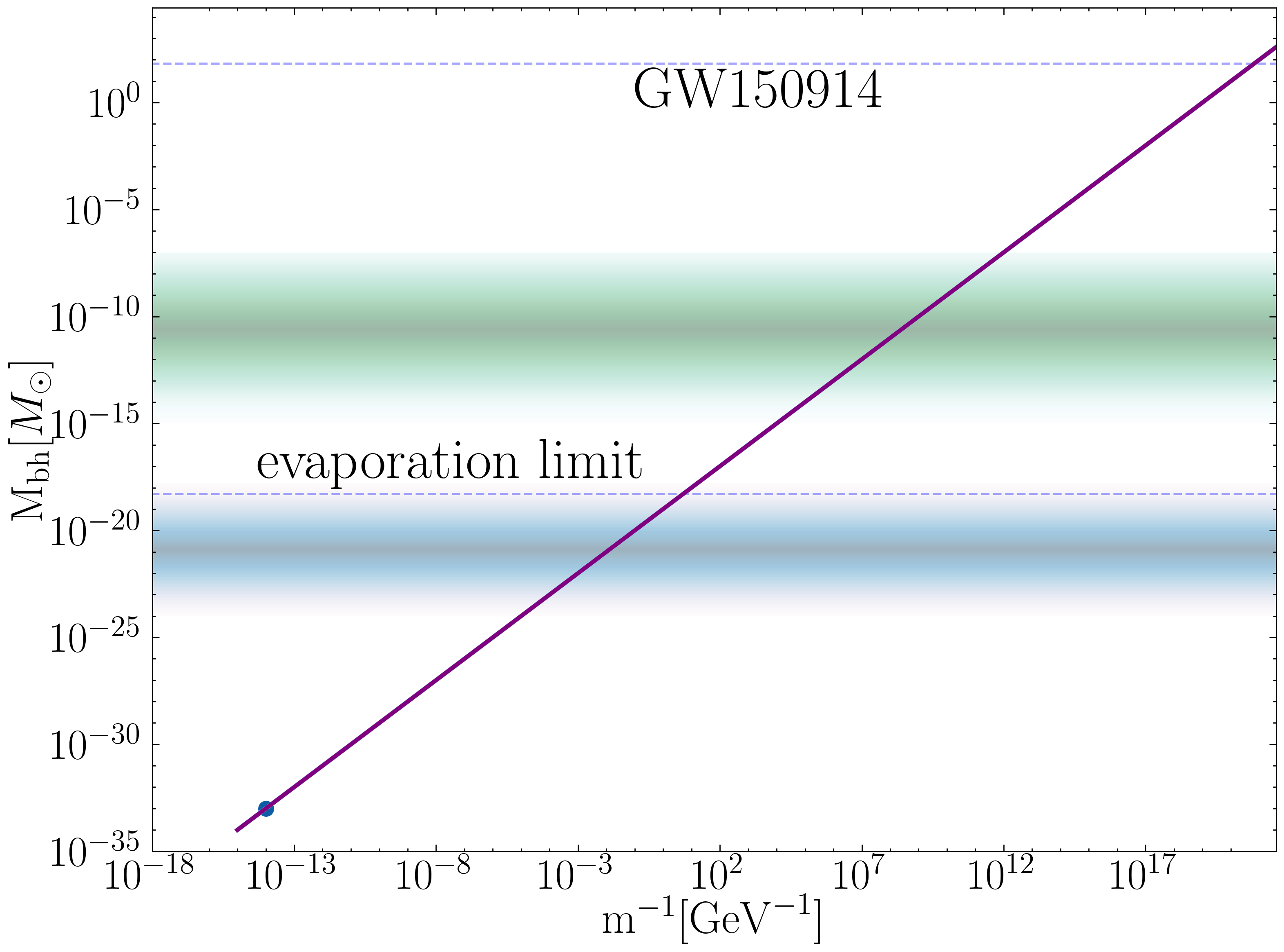}
        \caption{Scaling relation between the mass of the primordial black hole and the scalar mass $m$ for a fiducial model of $\alpha=0.18$ and $\beta=18$ (solid purple line). The ``evaporation limit'' line corresponds to the mass of a primordial black hole that would evaporate now due to the Hawking radiation, and the mass of the merged GW150914 black hole is also noted. The left bottom dot corresponds to the case of primordial black holes generated in oscillon preheating in the inflation model (\ref{eq:3}) subject to the condition (\ref{eq:7}). The cyan and green band correspond to the most sensitive frequency band of aLIGO/Virgo and LISA respectively, which may capture the stochastic gravitational wave background generated in oscillon preheating-like scenarios.}
        \label{fig:scaling}
\end{figure}

So far, we have focused on a preheating model where the $\phi$ field inflates the universe before rolling down to reheat, for which the $m$ parameter in the potential is fixed by the CMB observations, Eq.~(\ref{eq:7}). In this scenario, the produced black holes have relatively small masses and thus do not survive to the present day. However, in hybrid inflation or early universe models where another scalar field acquires a dominant VEV, the corresponding theory parameter $m$ is not subject to the restriction (\ref{eq:7}), and the ``preheating'' can happen at much low energy scales. In these scenarios, oscillons can collapse to much more massive black holes, which can survive to be primordial black holes in the current universe, with many interesting phenomenological applications (see e.g.~\cite{Bird:2016dcv, Sasaki:2016jop, Carr:2020xqk} and references therein). In particular, they may be an appreciable component of dark matter, the phenomenological implications of which is left for future work. Indeed, the simulations done in this paper are directly applicable for these models. To see this, simply note that action (\ref{action00}) is invariant under the scaling 
\be
x^\mu \to \li x^\mu,~~~~~V\to \li^{-2} V .
\ee
Alternatively, we can define dimensionless variables $\tilde x^\mu = m x^\mu,~\tilde\phi=\phi/M_{\rm pl}$ and then $M^2_{\rm pl} /m^2$ factors out of the action (\ref{action00}), leaving $\alpha$ and $\beta$ the only relevant dimensionless theory parameters in the model. When presenting our results previously, we have been essentially using $m$ as the units. In Fig.~\ref{fig:scaling}, we plot the scaling relation between the mass of the primordial black hole and the scalar mass $m$ in lower energy models where preheating-like scenarios occur. In oscillon preheating, a significant stochastic gravitational wave background can be generated during the formation of oscillons themselves, although when properly formed the oscillons do not emit much gravitational waves \cite{Zhou:2013tsa}. The frequencies of this stochastic background are very high for the inflation model (\ref{eq:3}) subject to the condition (\ref{eq:7}), but they can be much lower for other lower energy preheating-like models. In Fig.~\ref{fig:scaling}, we have also plotted the parameter ranges that fall within the most sensitive frequencies of the aLIGO and LISA experiments; see the cyan and green band. The detailed phenomenology of these models, particularly the implications for gravitational wave physics, is left for future work.

\section{Conclusion}
\label{sec:clu}

To summarize, we have investigated the effects of general relativity during the oscillon preheating scenario by comparing simulations in the full GR scheme with those in the FLRW scheme. The comparisons are done by looking at the fraction of energies contained in the oscillons as well as the power spectrum of the scalar field. By solving the Einstein equations exactly in the BSSN formalism, we have shown that the traditionally used FLRW simulation scheme is a good approximation when $\beta$ is large.  However, significant discrepancies do emerge when $\beta$ is small. In particular, for small $\beta$, simulations in the full GR scheme tend to condensate significantly more energies in oscillons and more quickly than the FLRW scheme, signaling stronger gravity effects in the formation of these oscillons. On the other hand, these strong gravity and non-perturbative effects make the full GR simulations with oscillons more difficult to converge. To increase the resolutions for the oscillons, we have utilized the AMR functionalities in our code. These AMR simulations indicate that oscillons are capable of collapsing into black holes in the preheating period for small $\beta$, confirming the strong self-gravity nature in the internal structure of these oscillons. We have also tentatively charted $\alpha$ and $\beta$ for the parameter region where black holes may form.

In the majority of the discussions in this paper, as a concrete and widely studied model, we have focused on the preheating scenario after inflation, for which the scalar mass $m$ is fixed by the CMB observations. However, as pointed out in Sec.~\ref{sec:largeBH}, we may choose to interpret $\phi$ as not the inflaton field but another field that develops a condensate later in the universe, for which case the oscillons and thus the black holes they collapse to can be much more massive, surviving in the present day as massive primordial black holes. If these primordial black holes are of appropriate astronomical scales, a merge event of a binary of them may produce gravitational wave signals observable in the aLIGO experiments. Also, the production of oscillons may then accompany a stochastic gravitational wave background in the frequency range of aLIGO or LISA. It would be interesting to investigate the phenomenological implications of these models in the future.

~\\
\noindent {\bf Acknowledgements}

We are grateful to Mustafa Amin, James B. Mertens, John T. Giblin, Jr and Glenn Starkman for helpful discussions. We would like to thank Yu-Xin Huang for helping test the code in the early stage of this work. This work made use of the High Performance Computing Resource in the Core Facility for Advanced Research Computing at Case Western Reserve University. CT were supported in part by grant DE-SC0009946 from the US DOE. SYZ acknowledges support from the starting grants from University of Science and Technology of China under grant No.~KY2030000089 and GG2030040375, and is also supported by National Natural Science Foundation of China under grant No.~11947301 and 12075233, and supported by the Fundamental Research Funds for the Central Universities under grant No.~WK2030000036.\\

\bibliography{reference}

\begin{thebibliography}{78}
\expandafter\ifx\csname natexlab\endcsname\relax\def\natexlab#1{#1}\fi
\expandafter\ifx\csname bibnamefont\endcsname\relax
  \def\bibnamefont#1{#1}\fi
\expandafter\ifx\csname bibfnamefont\endcsname\relax
  \def\bibfnamefont#1{#1}\fi
\expandafter\ifx\csname citenamefont\endcsname\relax
  \def\citenamefont#1{#1}\fi
\expandafter\ifx\csname url\endcsname\relax
  \def\url#1{\texttt{#1}}\fi
\expandafter\ifx\csname urlprefix\endcsname\relax\def\urlprefix{URL }\fi
\providecommand{\bibinfo}[2]{#2}
\providecommand{\eprint}[2][]{\url{#2}}

\bibitem[{\citenamefont{Bogolyubsky and Makhankov}(1976)}]{Bogolyubsky:1976yu}
\bibinfo{author}{\bibfnamefont{I.~L.} \bibnamefont{Bogolyubsky}}
  \bibnamefont{and} \bibinfo{author}{\bibfnamefont{V.~G.}
  \bibnamefont{Makhankov}}, \bibinfo{journal}{Pisma Zh. Eksp. Teor. Fiz.}
  \textbf{\bibinfo{volume}{24}}, \bibinfo{pages}{15} (\bibinfo{year}{1976}).

\bibitem[{\citenamefont{Gleiser}(1994)}]{Gleiser:1993pt}
\bibinfo{author}{\bibfnamefont{M.}~\bibnamefont{Gleiser}},
  \bibinfo{journal}{Phys. Rev.} \textbf{\bibinfo{volume}{D49}},
  \bibinfo{pages}{2978} (\bibinfo{year}{1994}), \eprint{hep-ph/9308279}.

\bibitem[{\citenamefont{Copeland et~al.}(1995)\citenamefont{Copeland, Gleiser,
  and Muller}}]{Copeland:1995fq}
\bibinfo{author}{\bibfnamefont{E.~J.} \bibnamefont{Copeland}},
  \bibinfo{author}{\bibfnamefont{M.}~\bibnamefont{Gleiser}}, \bibnamefont{and}
  \bibinfo{author}{\bibfnamefont{H.~R.} \bibnamefont{Muller}},
  \bibinfo{journal}{Phys. Rev.} \textbf{\bibinfo{volume}{D52}},
  \bibinfo{pages}{1920} (\bibinfo{year}{1995}), \eprint{hep-ph/9503217}.

\bibitem[{\citenamefont{Honda and Choptuik}(2002)}]{Honda:2001xg}
\bibinfo{author}{\bibfnamefont{E.~P.} \bibnamefont{Honda}} \bibnamefont{and}
  \bibinfo{author}{\bibfnamefont{M.~W.} \bibnamefont{Choptuik}},
  \bibinfo{journal}{Phys. Rev.} \textbf{\bibinfo{volume}{D65}},
  \bibinfo{pages}{084037} (\bibinfo{year}{2002}), \eprint{hep-ph/0110065}.

\bibitem[{\citenamefont{Hindmarsh and Salmi}(2006)}]{Hindmarsh:2006ur}
\bibinfo{author}{\bibfnamefont{M.}~\bibnamefont{Hindmarsh}} \bibnamefont{and}
  \bibinfo{author}{\bibfnamefont{P.}~\bibnamefont{Salmi}},
  \bibinfo{journal}{Phys. Rev.} \textbf{\bibinfo{volume}{D74}},
  \bibinfo{pages}{105005} (\bibinfo{year}{2006}), \eprint{hep-th/0606016}.

\bibitem[{\citenamefont{Fodor et~al.}(2006)\citenamefont{Fodor, Forgacs,
  Grandclement, and Racz}}]{Fodor:2006zs}
\bibinfo{author}{\bibfnamefont{G.}~\bibnamefont{Fodor}},
  \bibinfo{author}{\bibfnamefont{P.}~\bibnamefont{Forgacs}},
  \bibinfo{author}{\bibfnamefont{P.}~\bibnamefont{Grandclement}},
  \bibnamefont{and} \bibinfo{author}{\bibfnamefont{I.}~\bibnamefont{Racz}},
  \bibinfo{journal}{Phys. Rev.} \textbf{\bibinfo{volume}{D74}},
  \bibinfo{pages}{124003} (\bibinfo{year}{2006}), \eprint{hep-th/0609023}.

\bibitem[{\citenamefont{Saffin and Tranberg}(2007)}]{Saffin:2006yk}
\bibinfo{author}{\bibfnamefont{P.~M.} \bibnamefont{Saffin}} \bibnamefont{and}
  \bibinfo{author}{\bibfnamefont{A.}~\bibnamefont{Tranberg}},
  \bibinfo{journal}{JHEP} \textbf{\bibinfo{volume}{01}}, \bibinfo{pages}{030}
  (\bibinfo{year}{2007}), \eprint{hep-th/0610191}.

\bibitem[{\citenamefont{Hindmarsh and Salmi}(2008)}]{Hindmarsh:2007jb}
\bibinfo{author}{\bibfnamefont{M.}~\bibnamefont{Hindmarsh}} \bibnamefont{and}
  \bibinfo{author}{\bibfnamefont{P.}~\bibnamefont{Salmi}},
  \bibinfo{journal}{Phys. Rev.} \textbf{\bibinfo{volume}{D77}},
  \bibinfo{pages}{105025} (\bibinfo{year}{2008}), \eprint{0712.0614}.

\bibitem[{\citenamefont{Fodor et~al.}(2008)\citenamefont{Fodor, Forgacs,
  Horvath, and Lukacs}}]{Fodor:2008es}
\bibinfo{author}{\bibfnamefont{G.}~\bibnamefont{Fodor}},
  \bibinfo{author}{\bibfnamefont{P.}~\bibnamefont{Forgacs}},
  \bibinfo{author}{\bibfnamefont{Z.}~\bibnamefont{Horvath}}, \bibnamefont{and}
  \bibinfo{author}{\bibfnamefont{A.}~\bibnamefont{Lukacs}},
  \bibinfo{journal}{Phys. Rev.} \textbf{\bibinfo{volume}{D78}},
  \bibinfo{pages}{025003} (\bibinfo{year}{2008}), \eprint{0802.3525}.

\bibitem[{\citenamefont{Gleiser and Sicilia}(2008)}]{Gleiser:2008ty}
\bibinfo{author}{\bibfnamefont{M.}~\bibnamefont{Gleiser}} \bibnamefont{and}
  \bibinfo{author}{\bibfnamefont{D.}~\bibnamefont{Sicilia}},
  \bibinfo{journal}{Phys. Rev. Lett.} \textbf{\bibinfo{volume}{101}},
  \bibinfo{pages}{011602} (\bibinfo{year}{2008}), \eprint{0804.0791}.

\bibitem[{\citenamefont{Hertzberg}(2010)}]{Hertzberg:2010yz}
\bibinfo{author}{\bibfnamefont{M.~P.} \bibnamefont{Hertzberg}},
  \bibinfo{journal}{Phys. Rev.} \textbf{\bibinfo{volume}{D82}},
  \bibinfo{pages}{045022} (\bibinfo{year}{2010}), \eprint{1003.3459}.

\bibitem[{\citenamefont{Salmi and Hindmarsh}(2012)}]{Salmi:2012ta}
\bibinfo{author}{\bibfnamefont{P.}~\bibnamefont{Salmi}} \bibnamefont{and}
  \bibinfo{author}{\bibfnamefont{M.}~\bibnamefont{Hindmarsh}},
  \bibinfo{journal}{Phys. Rev.} \textbf{\bibinfo{volume}{D85}},
  \bibinfo{pages}{085033} (\bibinfo{year}{2012}), \eprint{1201.1934}.

\bibitem[{\citenamefont{Amin}(2013)}]{Amin:2013ika}
\bibinfo{author}{\bibfnamefont{M.~A.} \bibnamefont{Amin}},
  \bibinfo{journal}{Phys. Rev.} \textbf{\bibinfo{volume}{D87}},
  \bibinfo{pages}{123505} (\bibinfo{year}{2013}), \eprint{1303.1102}.

\bibitem[{\citenamefont{Copeland et~al.}(2014)\citenamefont{Copeland, Saffin,
  and Zhou}}]{Copeland:2014qra}
\bibinfo{author}{\bibfnamefont{E.~J.} \bibnamefont{Copeland}},
  \bibinfo{author}{\bibfnamefont{P.~M.} \bibnamefont{Saffin}},
  \bibnamefont{and} \bibinfo{author}{\bibfnamefont{S.-Y.} \bibnamefont{Zhou}},
  \bibinfo{journal}{Phys. Rev. Lett.} \textbf{\bibinfo{volume}{113}},
  \bibinfo{pages}{231603} (\bibinfo{year}{2014}), \eprint{1409.3232}.

\bibitem[{\citenamefont{Krippendorf et~al.}(2018)\citenamefont{Krippendorf,
  Muia, and Quevedo}}]{Krippendorf:2018tei}
\bibinfo{author}{\bibfnamefont{S.}~\bibnamefont{Krippendorf}},
  \bibinfo{author}{\bibfnamefont{F.}~\bibnamefont{Muia}}, \bibnamefont{and}
  \bibinfo{author}{\bibfnamefont{F.}~\bibnamefont{Quevedo}},
  \bibinfo{journal}{JHEP} \textbf{\bibinfo{volume}{08}}, \bibinfo{pages}{070}
  (\bibinfo{year}{2018}), \eprint{1806.04690}.

\bibitem[{\citenamefont{Amin and Mocz}(2019)}]{Amin:2019ums}
\bibinfo{author}{\bibfnamefont{M.~A.} \bibnamefont{Amin}} \bibnamefont{and}
  \bibinfo{author}{\bibfnamefont{P.}~\bibnamefont{Mocz}},
  \bibinfo{journal}{Phys. Rev.} \textbf{\bibinfo{volume}{D100}},
  \bibinfo{pages}{063507} (\bibinfo{year}{2019}), \eprint{1902.07261}.

\bibitem[{\citenamefont{Gleiser}(2007)}]{Gleiser:2006te}
\bibinfo{author}{\bibfnamefont{M.}~\bibnamefont{Gleiser}},
  \bibinfo{journal}{Int. J. Mod. Phys.} \textbf{\bibinfo{volume}{D16}},
  \bibinfo{pages}{219} (\bibinfo{year}{2007}), \eprint{hep-th/0602187}.

\bibitem[{\citenamefont{Graham and Stamatopoulos}(2006)}]{Graham:2006xs}
\bibinfo{author}{\bibfnamefont{N.}~\bibnamefont{Graham}} \bibnamefont{and}
  \bibinfo{author}{\bibfnamefont{N.}~\bibnamefont{Stamatopoulos}},
  \bibinfo{journal}{Phys. Lett.} \textbf{\bibinfo{volume}{B639}},
  \bibinfo{pages}{541} (\bibinfo{year}{2006}), \eprint{hep-th/0604134}.

\bibitem[{\citenamefont{Amin}(2010)}]{Amin:2010xe}
\bibinfo{author}{\bibfnamefont{M.~A.} \bibnamefont{Amin}}
  (\bibinfo{year}{2010}), \eprint{1006.3075}.

\bibitem[{\citenamefont{Amin and Shirokoff}(2010)}]{Amin:2010jq}
\bibinfo{author}{\bibfnamefont{M.~A.} \bibnamefont{Amin}} \bibnamefont{and}
  \bibinfo{author}{\bibfnamefont{D.}~\bibnamefont{Shirokoff}},
  \bibinfo{journal}{Phys. Rev.} \textbf{\bibinfo{volume}{D81}},
  \bibinfo{pages}{085045} (\bibinfo{year}{2010}), \eprint{1002.3380}.

\bibitem[{\citenamefont{Amin et~al.}(2012)\citenamefont{Amin, Easther, Finkel,
  Flauger, and Hertzberg}}]{Amin:2011hj}
\bibinfo{author}{\bibfnamefont{M.~A.} \bibnamefont{Amin}},
  \bibinfo{author}{\bibfnamefont{R.}~\bibnamefont{Easther}},
  \bibinfo{author}{\bibfnamefont{H.}~\bibnamefont{Finkel}},
  \bibinfo{author}{\bibfnamefont{R.}~\bibnamefont{Flauger}}, \bibnamefont{and}
  \bibinfo{author}{\bibfnamefont{M.~P.} \bibnamefont{Hertzberg}},
  \bibinfo{journal}{Phys. Rev. Lett.} \textbf{\bibinfo{volume}{108}},
  \bibinfo{pages}{241302} (\bibinfo{year}{2012}), \eprint{1106.3335}.

\bibitem[{\citenamefont{Amin et~al.}(2010)\citenamefont{Amin, Easther, and
  Finkel}}]{Amin:2010dc}
\bibinfo{author}{\bibfnamefont{M.~A.} \bibnamefont{Amin}},
  \bibinfo{author}{\bibfnamefont{R.}~\bibnamefont{Easther}}, \bibnamefont{and}
  \bibinfo{author}{\bibfnamefont{H.}~\bibnamefont{Finkel}},
  \bibinfo{journal}{JCAP} \textbf{\bibinfo{volume}{1012}}, \bibinfo{pages}{001}
  (\bibinfo{year}{2010}), \eprint{1009.2505}.

\bibitem[{\citenamefont{Broadhead and McDonald}(2005)}]{Broadhead:2005hn}
\bibinfo{author}{\bibfnamefont{M.}~\bibnamefont{Broadhead}} \bibnamefont{and}
  \bibinfo{author}{\bibfnamefont{J.}~\bibnamefont{McDonald}},
  \bibinfo{journal}{Phys. Rev.} \textbf{\bibinfo{volume}{D72}},
  \bibinfo{pages}{043519} (\bibinfo{year}{2005}), \eprint{hep-ph/0503081}.

\bibitem[{\citenamefont{Farhi et~al.}(2008)\citenamefont{Farhi, Graham, Guth,
  Iqbal, Rosales, and Stamatopoulos}}]{Farhi:2007wj}
\bibinfo{author}{\bibfnamefont{E.}~\bibnamefont{Farhi}},
  \bibinfo{author}{\bibfnamefont{N.}~\bibnamefont{Graham}},
  \bibinfo{author}{\bibfnamefont{A.~H.} \bibnamefont{Guth}},
  \bibinfo{author}{\bibfnamefont{N.}~\bibnamefont{Iqbal}},
  \bibinfo{author}{\bibfnamefont{R.~R.} \bibnamefont{Rosales}},
  \bibnamefont{and}
  \bibinfo{author}{\bibfnamefont{N.}~\bibnamefont{Stamatopoulos}},
  \bibinfo{journal}{Phys. Rev.} \textbf{\bibinfo{volume}{D77}},
  \bibinfo{pages}{085019} (\bibinfo{year}{2008}), \eprint{0712.3034}.

\bibitem[{\citenamefont{Gleiser et~al.}(2011)\citenamefont{Gleiser, Graham, and
  Stamatopoulos}}]{Gleiser:2011xj}
\bibinfo{author}{\bibfnamefont{M.}~\bibnamefont{Gleiser}},
  \bibinfo{author}{\bibfnamefont{N.}~\bibnamefont{Graham}}, \bibnamefont{and}
  \bibinfo{author}{\bibfnamefont{N.}~\bibnamefont{Stamatopoulos}},
  \bibinfo{journal}{Phys. Rev.} \textbf{\bibinfo{volume}{D83}},
  \bibinfo{pages}{096010} (\bibinfo{year}{2011}), \eprint{1103.1911}.

\bibitem[{\citenamefont{Zhou et~al.}(2013)\citenamefont{Zhou, Copeland,
  Easther, Finkel, Mou, and Saffin}}]{Zhou:2013tsa}
\bibinfo{author}{\bibfnamefont{S.-Y.} \bibnamefont{Zhou}},
  \bibinfo{author}{\bibfnamefont{E.~J.} \bibnamefont{Copeland}},
  \bibinfo{author}{\bibfnamefont{R.}~\bibnamefont{Easther}},
  \bibinfo{author}{\bibfnamefont{H.}~\bibnamefont{Finkel}},
  \bibinfo{author}{\bibfnamefont{Z.-G.} \bibnamefont{Mou}}, \bibnamefont{and}
  \bibinfo{author}{\bibfnamefont{P.~M.} \bibnamefont{Saffin}},
  \bibinfo{journal}{JHEP} \textbf{\bibinfo{volume}{10}}, \bibinfo{pages}{026}
  (\bibinfo{year}{2013}), \eprint{1304.6094}.

\bibitem[{\citenamefont{Antusch and Orani}(2016)}]{Antusch:2015ziz}
\bibinfo{author}{\bibfnamefont{S.}~\bibnamefont{Antusch}} \bibnamefont{and}
  \bibinfo{author}{\bibfnamefont{S.}~\bibnamefont{Orani}},
  \bibinfo{journal}{JCAP} \textbf{\bibinfo{volume}{1603}}, \bibinfo{pages}{026}
  (\bibinfo{year}{2016}), \eprint{1511.02336}.

\bibitem[{\citenamefont{Lozanov and Amin}(2019)}]{Lozanov:2019ylm}
\bibinfo{author}{\bibfnamefont{K.~D.} \bibnamefont{Lozanov}} \bibnamefont{and}
  \bibinfo{author}{\bibfnamefont{M.~A.} \bibnamefont{Amin}},
  \bibinfo{journal}{Phys. Rev. D} \textbf{\bibinfo{volume}{99}},
  \bibinfo{pages}{123504} (\bibinfo{year}{2019}), \eprint{1902.06736}.

\bibitem[{\citenamefont{Traschen and Brandenberger}(1990)}]{Traschen:1990sw}
\bibinfo{author}{\bibfnamefont{J.~H.} \bibnamefont{Traschen}} \bibnamefont{and}
  \bibinfo{author}{\bibfnamefont{R.~H.} \bibnamefont{Brandenberger}},
  \bibinfo{journal}{Phys. Rev. D} \textbf{\bibinfo{volume}{42}},
  \bibinfo{pages}{2491} (\bibinfo{year}{1990}).

\bibitem[{\citenamefont{Dolgov and Kirilova}(1989)}]{Dolgov:1989sy}
\bibinfo{author}{\bibfnamefont{A.~D.} \bibnamefont{Dolgov}} \bibnamefont{and}
  \bibinfo{author}{\bibfnamefont{D.~P.} \bibnamefont{Kirilova}},
  \bibinfo{journal}{Sov. J. Nucl. Phys.} \textbf{\bibinfo{volume}{50}},
  \bibinfo{pages}{1006} (\bibinfo{year}{1989}), \bibinfo{note}{[Yad.
  Fiz.50,1621(1989)]}.

\bibitem[{\citenamefont{Shtanov et~al.}(1995)\citenamefont{Shtanov, Traschen,
  and Brandenberger}}]{Shtanov:1994ce}
\bibinfo{author}{\bibfnamefont{Y.}~\bibnamefont{Shtanov}},
  \bibinfo{author}{\bibfnamefont{J.~H.} \bibnamefont{Traschen}},
  \bibnamefont{and} \bibinfo{author}{\bibfnamefont{R.~H.}
  \bibnamefont{Brandenberger}}, \bibinfo{journal}{Phys. Rev.}
  \textbf{\bibinfo{volume}{D51}}, \bibinfo{pages}{5438} (\bibinfo{year}{1995}),
  \eprint{hep-ph/9407247}.

\bibitem[{\citenamefont{Khlebnikov and Tkachev}(1996)}]{Khlebnikov:1996mc}
\bibinfo{author}{\bibfnamefont{S.~{\relax Yu}.} \bibnamefont{Khlebnikov}}
  \bibnamefont{and} \bibinfo{author}{\bibfnamefont{I.~I.}
  \bibnamefont{Tkachev}}, \bibinfo{journal}{Phys. Rev. Lett.}
  \textbf{\bibinfo{volume}{77}}, \bibinfo{pages}{219} (\bibinfo{year}{1996}),
  \eprint{hep-ph/9603378}.

\bibitem[{\citenamefont{Kofman et~al.}(1997)\citenamefont{Kofman, Linde, and
  Starobinsky}}]{Kofman:1997yn}
\bibinfo{author}{\bibfnamefont{L.}~\bibnamefont{Kofman}},
  \bibinfo{author}{\bibfnamefont{A.~D.} \bibnamefont{Linde}}, \bibnamefont{and}
  \bibinfo{author}{\bibfnamefont{A.~A.} \bibnamefont{Starobinsky}},
  \bibinfo{journal}{Phys. Rev.} \textbf{\bibinfo{volume}{D56}},
  \bibinfo{pages}{3258} (\bibinfo{year}{1997}), \eprint{hep-ph/9704452}.

\bibitem[{\citenamefont{Felder et~al.}(2001)\citenamefont{Felder,
  Garcia-Bellido, Greene, Kofman, Linde, and Tkachev}}]{Felder:2000hj}
\bibinfo{author}{\bibfnamefont{G.~N.} \bibnamefont{Felder}},
  \bibinfo{author}{\bibfnamefont{J.}~\bibnamefont{Garcia-Bellido}},
  \bibinfo{author}{\bibfnamefont{P.~B.} \bibnamefont{Greene}},
  \bibinfo{author}{\bibfnamefont{L.}~\bibnamefont{Kofman}},
  \bibinfo{author}{\bibfnamefont{A.~D.} \bibnamefont{Linde}}, \bibnamefont{and}
  \bibinfo{author}{\bibfnamefont{I.}~\bibnamefont{Tkachev}},
  \bibinfo{journal}{Phys. Rev. Lett.} \textbf{\bibinfo{volume}{87}},
  \bibinfo{pages}{011601} (\bibinfo{year}{2001}), \eprint{hep-ph/0012142}.

\bibitem[{\citenamefont{Antusch et~al.}(2017)\citenamefont{Antusch, Cefala, and
  Orani}}]{Antusch:2016con}
\bibinfo{author}{\bibfnamefont{S.}~\bibnamefont{Antusch}},
  \bibinfo{author}{\bibfnamefont{F.}~\bibnamefont{Cefala}}, \bibnamefont{and}
  \bibinfo{author}{\bibfnamefont{S.}~\bibnamefont{Orani}},
  \bibinfo{journal}{Phys. Rev. Lett.} \textbf{\bibinfo{volume}{118}},
  \bibinfo{pages}{011303} (\bibinfo{year}{2017}), \bibinfo{note}{[Erratum:
  Phys. Rev. Lett.120,no.21,219901(2018)]}, \eprint{1607.01314}.

\bibitem[{\citenamefont{Liu et~al.}(2018)\citenamefont{Liu, Guo, Cai, and
  Shiu}}]{Liu:2017hua}
\bibinfo{author}{\bibfnamefont{J.}~\bibnamefont{Liu}},
  \bibinfo{author}{\bibfnamefont{Z.-K.} \bibnamefont{Guo}},
  \bibinfo{author}{\bibfnamefont{R.-G.} \bibnamefont{Cai}}, \bibnamefont{and}
  \bibinfo{author}{\bibfnamefont{G.}~\bibnamefont{Shiu}},
  \bibinfo{journal}{Phys. Rev. Lett.} \textbf{\bibinfo{volume}{120}},
  \bibinfo{pages}{031301} (\bibinfo{year}{2018}), \eprint{1707.09841}.

\bibitem[{\citenamefont{Antusch
  et~al.}(2018{\natexlab{a}})\citenamefont{Antusch, Cefala, Krippendorf, Muia,
  Orani, and Quevedo}}]{Antusch:2017flz}
\bibinfo{author}{\bibfnamefont{S.}~\bibnamefont{Antusch}},
  \bibinfo{author}{\bibfnamefont{F.}~\bibnamefont{Cefala}},
  \bibinfo{author}{\bibfnamefont{S.}~\bibnamefont{Krippendorf}},
  \bibinfo{author}{\bibfnamefont{F.}~\bibnamefont{Muia}},
  \bibinfo{author}{\bibfnamefont{S.}~\bibnamefont{Orani}}, \bibnamefont{and}
  \bibinfo{author}{\bibfnamefont{F.}~\bibnamefont{Quevedo}},
  \bibinfo{journal}{JHEP} \textbf{\bibinfo{volume}{01}}, \bibinfo{pages}{083}
  (\bibinfo{year}{2018}{\natexlab{a}}), \eprint{1708.08922}.

\bibitem[{\citenamefont{Antusch
  et~al.}(2018{\natexlab{b}})\citenamefont{Antusch, Cefala, and
  Orani}}]{Antusch:2017vga}
\bibinfo{author}{\bibfnamefont{S.}~\bibnamefont{Antusch}},
  \bibinfo{author}{\bibfnamefont{F.}~\bibnamefont{Cefala}}, \bibnamefont{and}
  \bibinfo{author}{\bibfnamefont{S.}~\bibnamefont{Orani}},
  \bibinfo{journal}{JCAP} \textbf{\bibinfo{volume}{1803}}, \bibinfo{pages}{032}
  (\bibinfo{year}{2018}{\natexlab{b}}), \eprint{1712.03231}.

\bibitem[{\citenamefont{Zhou}(2015)}]{Zhou:2015yfa}
\bibinfo{author}{\bibfnamefont{S.-Y.} \bibnamefont{Zhou}},
  \bibinfo{journal}{JCAP} \textbf{\bibinfo{volume}{1506}}, \bibinfo{pages}{033}
  (\bibinfo{year}{2015}), \eprint{1501.01217}.

\bibitem[{\citenamefont{Amin et~al.}(2018)\citenamefont{Amin, Braden, Copeland,
  Giblin, Solorio, Weiner, and Zhou}}]{Amin:2018xfe}
\bibinfo{author}{\bibfnamefont{M.~A.} \bibnamefont{Amin}},
  \bibinfo{author}{\bibfnamefont{J.}~\bibnamefont{Braden}},
  \bibinfo{author}{\bibfnamefont{E.~J.} \bibnamefont{Copeland}},
  \bibinfo{author}{\bibfnamefont{J.~T.} \bibnamefont{Giblin}},
  \bibinfo{author}{\bibfnamefont{C.}~\bibnamefont{Solorio}},
  \bibinfo{author}{\bibfnamefont{Z.~J.} \bibnamefont{Weiner}},
  \bibnamefont{and} \bibinfo{author}{\bibfnamefont{S.-Y.} \bibnamefont{Zhou}},
  \bibinfo{journal}{Phys. Rev.} \textbf{\bibinfo{volume}{D98}},
  \bibinfo{pages}{024040} (\bibinfo{year}{2018}), \eprint{1803.08047}.

\bibitem[{\citenamefont{Sang and Huang}(2019)}]{Sang:2019ndv}
\bibinfo{author}{\bibfnamefont{Y.}~\bibnamefont{Sang}} \bibnamefont{and}
  \bibinfo{author}{\bibfnamefont{Q.-G.} \bibnamefont{Huang}},
  \bibinfo{journal}{Phys. Rev.} \textbf{\bibinfo{volume}{D100}},
  \bibinfo{pages}{063516} (\bibinfo{year}{2019}), \eprint{1905.00371}.

\bibitem[{\citenamefont{Aghanim et~al.}(2018)}]{Aghanim:2018eyx}
\bibinfo{author}{\bibfnamefont{N.}~\bibnamefont{Aghanim}} \bibnamefont{et~al.}
  (\bibinfo{collaboration}{Planck}) (\bibinfo{year}{2018}),
  \eprint{1807.06209}.

\bibitem[{\citenamefont{East et~al.}(2016)\citenamefont{East, Kleban, Linde,
  and Senatore}}]{East:2015ggf}
\bibinfo{author}{\bibfnamefont{W.~E.} \bibnamefont{East}},
  \bibinfo{author}{\bibfnamefont{M.}~\bibnamefont{Kleban}},
  \bibinfo{author}{\bibfnamefont{A.}~\bibnamefont{Linde}}, \bibnamefont{and}
  \bibinfo{author}{\bibfnamefont{L.}~\bibnamefont{Senatore}},
  \bibinfo{journal}{JCAP} \textbf{\bibinfo{volume}{1609}}, \bibinfo{pages}{010}
  (\bibinfo{year}{2016}), \eprint{1511.05143}.

\bibitem[{\citenamefont{East et~al.}(2017)\citenamefont{East, Kearney, Shakya,
  Yoo, and Zurek}}]{East:2016anr}
\bibinfo{author}{\bibfnamefont{W.~E.} \bibnamefont{East}},
  \bibinfo{author}{\bibfnamefont{J.}~\bibnamefont{Kearney}},
  \bibinfo{author}{\bibfnamefont{B.}~\bibnamefont{Shakya}},
  \bibinfo{author}{\bibfnamefont{H.}~\bibnamefont{Yoo}}, \bibnamefont{and}
  \bibinfo{author}{\bibfnamefont{K.~M.} \bibnamefont{Zurek}},
  \bibinfo{journal}{Phys. Rev.} \textbf{\bibinfo{volume}{D95}},
  \bibinfo{pages}{023526} (\bibinfo{year}{2017}), \bibinfo{note}{[Phys.
  Rev.D95,023526(2017)]}, \eprint{1607.00381}.

\bibitem[{\citenamefont{Clough et~al.}(2018)\citenamefont{Clough, Flauger, and
  Lim}}]{Clough:2017efm}
\bibinfo{author}{\bibfnamefont{K.}~\bibnamefont{Clough}},
  \bibinfo{author}{\bibfnamefont{R.}~\bibnamefont{Flauger}}, \bibnamefont{and}
  \bibinfo{author}{\bibfnamefont{E.~A.} \bibnamefont{Lim}},
  \bibinfo{journal}{JCAP} \textbf{\bibinfo{volume}{1805}}, \bibinfo{pages}{065}
  (\bibinfo{year}{2018}), \eprint{1712.07352}.

\bibitem[{\citenamefont{Bloomfield et~al.}(2019)\citenamefont{Bloomfield,
  Fitzpatrick, Hilbert, and Kaiser}}]{Bloomfield:2019rbs}
\bibinfo{author}{\bibfnamefont{J.~K.} \bibnamefont{Bloomfield}},
  \bibinfo{author}{\bibfnamefont{P.}~\bibnamefont{Fitzpatrick}},
  \bibinfo{author}{\bibfnamefont{K.}~\bibnamefont{Hilbert}}, \bibnamefont{and}
  \bibinfo{author}{\bibfnamefont{D.~I.} \bibnamefont{Kaiser}},
  \bibinfo{journal}{Phys. Rev.} \textbf{\bibinfo{volume}{D100}},
  \bibinfo{pages}{063512} (\bibinfo{year}{2019}), \eprint{1906.08651}.

\bibitem[{\citenamefont{Giblin et~al.}(2016{\natexlab{a}})\citenamefont{Giblin,
  Mertens, and Starkman}}]{Giblin:2015vwq}
\bibinfo{author}{\bibfnamefont{J.~T.} \bibnamefont{Giblin}},
  \bibinfo{author}{\bibfnamefont{J.~B.} \bibnamefont{Mertens}},
  \bibnamefont{and} \bibinfo{author}{\bibfnamefont{G.~D.}
  \bibnamefont{Starkman}}, \bibinfo{journal}{Phys. Rev. Lett.}
  \textbf{\bibinfo{volume}{116}}, \bibinfo{pages}{251301}
  (\bibinfo{year}{2016}{\natexlab{a}}), \eprint{1511.01105}.

\bibitem[{\citenamefont{Bentivegna and Bruni}(2016)}]{Bentivegna:2015flc}
\bibinfo{author}{\bibfnamefont{E.}~\bibnamefont{Bentivegna}} \bibnamefont{and}
  \bibinfo{author}{\bibfnamefont{M.}~\bibnamefont{Bruni}},
  \bibinfo{journal}{Phys. Rev. Lett.} \textbf{\bibinfo{volume}{116}},
  \bibinfo{pages}{251302} (\bibinfo{year}{2016}), \eprint{1511.05124}.

\bibitem[{\citenamefont{Giblin et~al.}(2016{\natexlab{b}})\citenamefont{Giblin,
  Mertens, and Starkman}}]{Giblin:2016mjp}
\bibinfo{author}{\bibfnamefont{J.~T.} \bibnamefont{Giblin}},
  \bibinfo{author}{\bibfnamefont{J.~B.} \bibnamefont{Mertens}},
  \bibnamefont{and} \bibinfo{author}{\bibfnamefont{G.~D.}
  \bibnamefont{Starkman}}, \bibinfo{journal}{Astrophys. J.}
  \textbf{\bibinfo{volume}{833}}, \bibinfo{pages}{247}
  (\bibinfo{year}{2016}{\natexlab{b}}), \eprint{1608.04403}.

\bibitem[{\citenamefont{Macpherson et~al.}(2017)\citenamefont{Macpherson,
  Lasky, and Price}}]{Macpherson:2016ict}
\bibinfo{author}{\bibfnamefont{H.~J.} \bibnamefont{Macpherson}},
  \bibinfo{author}{\bibfnamefont{P.~D.} \bibnamefont{Lasky}}, \bibnamefont{and}
  \bibinfo{author}{\bibfnamefont{D.~J.} \bibnamefont{Price}},
  \bibinfo{journal}{Phys. Rev.} \textbf{\bibinfo{volume}{D95}},
  \bibinfo{pages}{064028} (\bibinfo{year}{2017}), \eprint{1611.05447}.

\bibitem[{\citenamefont{Giblin et~al.}(2017)\citenamefont{Giblin, Mertens, and
  Starkman}}]{Giblin:2017juu}
\bibinfo{author}{\bibfnamefont{J.~T.} \bibnamefont{Giblin}},
  \bibinfo{author}{\bibfnamefont{J.~B.} \bibnamefont{Mertens}},
  \bibnamefont{and} \bibinfo{author}{\bibfnamefont{G.~D.}
  \bibnamefont{Starkman}}, \bibinfo{journal}{Class. Quant. Grav.}
  \textbf{\bibinfo{volume}{34}}, \bibinfo{pages}{214001}
  (\bibinfo{year}{2017}), \eprint{1704.04307}.

\bibitem[{\citenamefont{East et~al.}(2018)\citenamefont{East, Wojtak, and
  Abel}}]{East:2017qmk}
\bibinfo{author}{\bibfnamefont{W.~E.} \bibnamefont{East}},
  \bibinfo{author}{\bibfnamefont{R.}~\bibnamefont{Wojtak}}, \bibnamefont{and}
  \bibinfo{author}{\bibfnamefont{T.}~\bibnamefont{Abel}},
  \bibinfo{journal}{Phys. Rev.} \textbf{\bibinfo{volume}{D97}},
  \bibinfo{pages}{043509} (\bibinfo{year}{2018}), \eprint{1711.06681}.

\bibitem[{\citenamefont{Wang}(2018)}]{Wang:2018qfr}
\bibinfo{author}{\bibfnamefont{K.}~\bibnamefont{Wang}}, \bibinfo{journal}{Eur.
  Phys. J.} \textbf{\bibinfo{volume}{C78}}, \bibinfo{pages}{629}
  (\bibinfo{year}{2018}), \eprint{1801.08362}.

\bibitem[{\citenamefont{Macpherson et~al.}(2018)\citenamefont{Macpherson,
  Lasky, and Price}}]{Macpherson:2018akp}
\bibinfo{author}{\bibfnamefont{H.~J.} \bibnamefont{Macpherson}},
  \bibinfo{author}{\bibfnamefont{P.~D.} \bibnamefont{Lasky}}, \bibnamefont{and}
  \bibinfo{author}{\bibfnamefont{D.~J.} \bibnamefont{Price}},
  \bibinfo{journal}{Astrophys. J.} \textbf{\bibinfo{volume}{865}},
  \bibinfo{pages}{L4} (\bibinfo{year}{2018}), \eprint{1807.01714}.

\bibitem[{\citenamefont{Macpherson et~al.}(2019)\citenamefont{Macpherson,
  Price, and Lasky}}]{Macpherson:2018btl}
\bibinfo{author}{\bibfnamefont{H.~J.} \bibnamefont{Macpherson}},
  \bibinfo{author}{\bibfnamefont{D.~J.} \bibnamefont{Price}}, \bibnamefont{and}
  \bibinfo{author}{\bibfnamefont{P.~D.} \bibnamefont{Lasky}},
  \bibinfo{journal}{Phys. Rev.} \textbf{\bibinfo{volume}{D99}},
  \bibinfo{pages}{063522} (\bibinfo{year}{2019}), \eprint{1807.01711}.

\bibitem[{\citenamefont{Giblin et~al.}(2019{\natexlab{a}})\citenamefont{Giblin,
  Mertens, Starkman, and Tian}}]{Giblin:2018ndw}
\bibinfo{author}{\bibfnamefont{J.~T.} \bibnamefont{Giblin}},
  \bibinfo{author}{\bibfnamefont{J.~B.} \bibnamefont{Mertens}},
  \bibinfo{author}{\bibfnamefont{G.~D.} \bibnamefont{Starkman}},
  \bibnamefont{and} \bibinfo{author}{\bibfnamefont{C.}~\bibnamefont{Tian}},
  \bibinfo{journal}{Phys. Rev.} \textbf{\bibinfo{volume}{D99}},
  \bibinfo{pages}{023527} (\bibinfo{year}{2019}{\natexlab{a}}),
  \eprint{1810.05203}.

\bibitem[{\citenamefont{Muia et~al.}(2019)\citenamefont{Muia, Cicoli, Clough,
  Pedro, Quevedo, and Vacca}}]{Muia:2019coe}
\bibinfo{author}{\bibfnamefont{F.}~\bibnamefont{Muia}},
  \bibinfo{author}{\bibfnamefont{M.}~\bibnamefont{Cicoli}},
  \bibinfo{author}{\bibfnamefont{K.}~\bibnamefont{Clough}},
  \bibinfo{author}{\bibfnamefont{F.}~\bibnamefont{Pedro}},
  \bibinfo{author}{\bibfnamefont{F.}~\bibnamefont{Quevedo}}, \bibnamefont{and}
  \bibinfo{author}{\bibfnamefont{G.~P.} \bibnamefont{Vacca}},
  \bibinfo{journal}{JCAP} \textbf{\bibinfo{volume}{1907}}, \bibinfo{pages}{044}
  (\bibinfo{year}{2019}), \eprint{1906.09346}.

\bibitem[{\citenamefont{Giblin and Tishue}(2019)}]{Giblin:2019nuv}
\bibinfo{author}{\bibfnamefont{J.~T.} \bibnamefont{Giblin}} \bibnamefont{and}
  \bibinfo{author}{\bibfnamefont{A.~J.} \bibnamefont{Tishue}},
  \bibinfo{journal}{Phys. Rev.} \textbf{\bibinfo{volume}{D100}},
  \bibinfo{pages}{063543} (\bibinfo{year}{2019}), \eprint{1907.10601}.

\bibitem[{\citenamefont{Coleman}(1985)}]{Coleman:1985ki}
\bibinfo{author}{\bibfnamefont{S.~R.} \bibnamefont{Coleman}},
  \bibinfo{journal}{Nucl. Phys. B} \textbf{\bibinfo{volume}{262}},
  \bibinfo{pages}{263} (\bibinfo{year}{1985}), \bibinfo{note}{[\textit{Erratum
  Nucl.Phys.B} \textbf{269} (1986) 744]}.

\bibitem[{\citenamefont{Andersen and Tranberg}(2012)}]{Andersen:2012wg}
\bibinfo{author}{\bibfnamefont{E.~A.} \bibnamefont{Andersen}} \bibnamefont{and}
  \bibinfo{author}{\bibfnamefont{A.}~\bibnamefont{Tranberg}},
  \bibinfo{journal}{JHEP} \textbf{\bibinfo{volume}{12}}, \bibinfo{pages}{016}
  (\bibinfo{year}{2012}), \eprint{1210.2227}.

\bibitem[{\citenamefont{Dolgov and Kirilova}(1990)}]{Dolgov:1989us}
\bibinfo{author}{\bibfnamefont{A.}~\bibnamefont{Dolgov}} \bibnamefont{and}
  \bibinfo{author}{\bibfnamefont{D.}~\bibnamefont{Kirilova}},
  \bibinfo{journal}{Sov. J. Nucl. Phys.} \textbf{\bibinfo{volume}{51}},
  \bibinfo{pages}{172} (\bibinfo{year}{1990}).

\bibitem[{\citenamefont{Silverstein and Westphal}(2008)}]{Silverstein:2008sg}
\bibinfo{author}{\bibfnamefont{E.}~\bibnamefont{Silverstein}} \bibnamefont{and}
  \bibinfo{author}{\bibfnamefont{A.}~\bibnamefont{Westphal}},
  \bibinfo{journal}{Phys. Rev.} \textbf{\bibinfo{volume}{D78}},
  \bibinfo{pages}{106003} (\bibinfo{year}{2008}), \eprint{0803.3085}.

\bibitem[{\citenamefont{McAllister et~al.}(2010)\citenamefont{McAllister,
  Silverstein, and Westphal}}]{McAllister:2008hb}
\bibinfo{author}{\bibfnamefont{L.}~\bibnamefont{McAllister}},
  \bibinfo{author}{\bibfnamefont{E.}~\bibnamefont{Silverstein}},
  \bibnamefont{and} \bibinfo{author}{\bibfnamefont{A.}~\bibnamefont{Westphal}},
  \bibinfo{journal}{Phys. Rev.} \textbf{\bibinfo{volume}{D82}},
  \bibinfo{pages}{046003} (\bibinfo{year}{2010}), \eprint{0808.0706}.

\bibitem[{\citenamefont{Dong et~al.}(2011)\citenamefont{Dong, Horn,
  Silverstein, and Westphal}}]{Dong:2010in}
\bibinfo{author}{\bibfnamefont{X.}~\bibnamefont{Dong}},
  \bibinfo{author}{\bibfnamefont{B.}~\bibnamefont{Horn}},
  \bibinfo{author}{\bibfnamefont{E.}~\bibnamefont{Silverstein}},
  \bibnamefont{and} \bibinfo{author}{\bibfnamefont{A.}~\bibnamefont{Westphal}},
  \bibinfo{journal}{Phys. Rev.} \textbf{\bibinfo{volume}{D84}},
  \bibinfo{pages}{026011} (\bibinfo{year}{2011}), \eprint{1011.4521}.

\bibitem[{\citenamefont{Mertens et~al.}(2016)\citenamefont{Mertens, Giblin, and
  Starkman}}]{Mertens:2015ttp}
\bibinfo{author}{\bibfnamefont{J.~B.} \bibnamefont{Mertens}},
  \bibinfo{author}{\bibfnamefont{J.~T.} \bibnamefont{Giblin}},
  \bibnamefont{and} \bibinfo{author}{\bibfnamefont{G.~D.}
  \bibnamefont{Starkman}}, \bibinfo{journal}{Phys. Rev.}
  \textbf{\bibinfo{volume}{D93}}, \bibinfo{pages}{124059}
  (\bibinfo{year}{2016}), \eprint{1511.01106}.

\bibitem[{\citenamefont{Nakamura et~al.}(1987)\citenamefont{Nakamura, Oohara,
  and Kojima}}]{Nakamura:1987zz}
\bibinfo{author}{\bibfnamefont{T.}~\bibnamefont{Nakamura}},
  \bibinfo{author}{\bibfnamefont{K.}~\bibnamefont{Oohara}}, \bibnamefont{and}
  \bibinfo{author}{\bibfnamefont{Y.}~\bibnamefont{Kojima}},
  \bibinfo{journal}{Prog. Theor. Phys. Suppl.} \textbf{\bibinfo{volume}{90}},
  \bibinfo{pages}{1} (\bibinfo{year}{1987}).

\bibitem[{\citenamefont{Shibata and Nakamura}(1995)}]{Shibata:1995we}
\bibinfo{author}{\bibfnamefont{M.}~\bibnamefont{Shibata}} \bibnamefont{and}
  \bibinfo{author}{\bibfnamefont{T.}~\bibnamefont{Nakamura}},
  \bibinfo{journal}{Phys. Rev.} \textbf{\bibinfo{volume}{D52}},
  \bibinfo{pages}{5428} (\bibinfo{year}{1995}).

\bibitem[{\citenamefont{Baumgarte and Shapiro}(1998)}]{Baumgarte:1998te}
\bibinfo{author}{\bibfnamefont{T.~W.} \bibnamefont{Baumgarte}}
  \bibnamefont{and} \bibinfo{author}{\bibfnamefont{S.~L.}
  \bibnamefont{Shapiro}}, \bibinfo{journal}{Phys. Rev.}
  \textbf{\bibinfo{volume}{D59}}, \bibinfo{pages}{024007}
  (\bibinfo{year}{1998}), \eprint{gr-qc/9810065}.

\bibitem[{\citenamefont{Giblin et~al.}(2019{\natexlab{b}})\citenamefont{Giblin,
  Mertens, Starkman, and Tian}}]{Giblin:2019pql}
\bibinfo{author}{\bibfnamefont{J.~T.} \bibnamefont{Giblin}},
  \bibinfo{author}{\bibfnamefont{J.~B.} \bibnamefont{Mertens}},
  \bibinfo{author}{\bibfnamefont{G.~D.} \bibnamefont{Starkman}},
  \bibnamefont{and} \bibinfo{author}{\bibfnamefont{C.}~\bibnamefont{Tian}},
  \bibinfo{journal}{Class. Quant. Grav.} \textbf{\bibinfo{volume}{36}},
  \bibinfo{pages}{195009} (\bibinfo{year}{2019}{\natexlab{b}}),
  \eprint{1903.01490}.

\bibitem[{\citenamefont{Wissink et~al.}(2001)\citenamefont{Wissink, Hornung,
  Kohn, Smith, and Elliott}}]{DBLP:conf/sc/WissinkHKSE01}
\bibinfo{author}{\bibfnamefont{A.~M.} \bibnamefont{Wissink}},
  \bibinfo{author}{\bibfnamefont{R.~D.} \bibnamefont{Hornung}},
  \bibinfo{author}{\bibfnamefont{S.~R.} \bibnamefont{Kohn}},
  \bibinfo{author}{\bibfnamefont{S.~S.} \bibnamefont{Smith}}, \bibnamefont{and}
  \bibinfo{author}{\bibfnamefont{N.}~\bibnamefont{Elliott}}, in
  \emph{\bibinfo{booktitle}{Proceedings of the 2001 {ACM/IEEE} conference on
  Supercomputing, Denver, CO, USA, November 10-16, 2001, {CD-ROM}}}
  (\bibinfo{year}{2001}), p.~\bibinfo{pages}{6},
  \urlprefix\url{http://doi.acm.org/10.1145/582034.582040}.

\bibitem[{\citenamefont{Gunney and Anderson}(2016)}]{GUNNEY201665}
\bibinfo{author}{\bibfnamefont{B.~T.} \bibnamefont{Gunney}} \bibnamefont{and}
  \bibinfo{author}{\bibfnamefont{R.~W.} \bibnamefont{Anderson}},
  \bibinfo{journal}{Journal of Parallel and Distributed Computing}
  \textbf{\bibinfo{volume}{89}}, \bibinfo{pages}{65 } (\bibinfo{year}{2016}),
  ISSN \bibinfo{issn}{0743-7315},
  \urlprefix\url{http://www.sciencedirect.com/science/article/pii/S0743731515002129}.

\bibitem[{\citenamefont{Thornburg}(2004)}]{Thornburg:2003sf}
\bibinfo{author}{\bibfnamefont{J.}~\bibnamefont{Thornburg}},
  \bibinfo{journal}{Class. Quant. Grav.} \textbf{\bibinfo{volume}{21}},
  \bibinfo{pages}{743} (\bibinfo{year}{2004}), \eprint{gr-qc/0306056}.

\bibitem[{\citenamefont{Press et~al.}(2003)\citenamefont{Press, Teukolsky,
  Vettering, and Flannery}}]{Press_2003}
\bibinfo{author}{\bibfnamefont{W.~H.} \bibnamefont{Press}},
  \bibinfo{author}{\bibfnamefont{S.~A.} \bibnamefont{Teukolsky}},
  \bibinfo{author}{\bibfnamefont{W.~T.} \bibnamefont{Vettering}},
  \bibnamefont{and} \bibinfo{author}{\bibfnamefont{B.~P.}
  \bibnamefont{Flannery}}, \bibinfo{journal}{European Journal of Physics}
  \textbf{\bibinfo{volume}{24}} (\bibinfo{year}{2003}).

\bibitem[{\citenamefont{Brown et~al.}(2007)\citenamefont{Brown, Sarbach,
  Schnetter, Tiglio, Diener, Hawke, and Pollney}}]{Brown:2007pg}
\bibinfo{author}{\bibfnamefont{J.~D.} \bibnamefont{Brown}},
  \bibinfo{author}{\bibfnamefont{O.}~\bibnamefont{Sarbach}},
  \bibinfo{author}{\bibfnamefont{E.}~\bibnamefont{Schnetter}},
  \bibinfo{author}{\bibfnamefont{M.}~\bibnamefont{Tiglio}},
  \bibinfo{author}{\bibfnamefont{P.}~\bibnamefont{Diener}},
  \bibinfo{author}{\bibfnamefont{I.}~\bibnamefont{Hawke}}, \bibnamefont{and}
  \bibinfo{author}{\bibfnamefont{D.}~\bibnamefont{Pollney}},
  \bibinfo{journal}{Phys. Rev.} \textbf{\bibinfo{volume}{D76}},
  \bibinfo{pages}{081503} (\bibinfo{year}{2007}), \eprint{0707.3101}.

\bibitem[{\citenamefont{Oberkampf and Roy}(2010)}]{oberkampf2010verification}
\bibinfo{author}{\bibfnamefont{W.~L.} \bibnamefont{Oberkampf}}
  \bibnamefont{and} \bibinfo{author}{\bibfnamefont{C.~J.} \bibnamefont{Roy}},
  \emph{\bibinfo{title}{Verification and validation in scientific computing}}
  (\bibinfo{publisher}{Cambridge University Press}, \bibinfo{year}{2010}).

\bibitem[{\citenamefont{Bird et~al.}(2016)\citenamefont{Bird, Cholis, Muñoz,
  Ali-Haïmoud, Kamionkowski, Kovetz, Raccanelli, and Riess}}]{Bird:2016dcv}
\bibinfo{author}{\bibfnamefont{S.}~\bibnamefont{Bird}},
  \bibinfo{author}{\bibfnamefont{I.}~\bibnamefont{Cholis}},
  \bibinfo{author}{\bibfnamefont{J.~B.} \bibnamefont{Muñoz}},
  \bibinfo{author}{\bibfnamefont{Y.}~\bibnamefont{Ali-Haïmoud}},
  \bibinfo{author}{\bibfnamefont{M.}~\bibnamefont{Kamionkowski}},
  \bibinfo{author}{\bibfnamefont{E.~D.} \bibnamefont{Kovetz}},
  \bibinfo{author}{\bibfnamefont{A.}~\bibnamefont{Raccanelli}},
  \bibnamefont{and} \bibinfo{author}{\bibfnamefont{A.~G.} \bibnamefont{Riess}},
  \bibinfo{journal}{Phys. Rev. Lett.} \textbf{\bibinfo{volume}{116}},
  \bibinfo{pages}{201301} (\bibinfo{year}{2016}), \eprint{1603.00464}.

\bibitem[{\citenamefont{Sasaki et~al.}(2016)\citenamefont{Sasaki, Suyama,
  Tanaka, and Yokoyama}}]{Sasaki:2016jop}
\bibinfo{author}{\bibfnamefont{M.}~\bibnamefont{Sasaki}},
  \bibinfo{author}{\bibfnamefont{T.}~\bibnamefont{Suyama}},
  \bibinfo{author}{\bibfnamefont{T.}~\bibnamefont{Tanaka}}, \bibnamefont{and}
  \bibinfo{author}{\bibfnamefont{S.}~\bibnamefont{Yokoyama}},
  \bibinfo{journal}{Phys. Rev. Lett.} \textbf{\bibinfo{volume}{117}},
  \bibinfo{pages}{061101} (\bibinfo{year}{2016}), \bibinfo{note}{[Erratum:
  Phys.Rev.Lett. 121, 059901 (2018)]}, \eprint{1603.08338}.

\bibitem[{\citenamefont{Carr and Kuhnel}(2020)}]{Carr:2020xqk}
\bibinfo{author}{\bibfnamefont{B.}~\bibnamefont{Carr}} \bibnamefont{and}
  \bibinfo{author}{\bibfnamefont{F.}~\bibnamefont{Kuhnel}}
  (\bibinfo{year}{2020}), \eprint{2006.02838}.

\end{thebibliography}

\end{document}